\documentclass[manuscript]{aastex}

\shorttitle{Common Proper Motion Stars}
\shortauthors{Janes}

\begin{document}


\title{Rotation Periods of Wide Binaries  \\
    in the Kepler Field}

\author{K.A.Janes}
\affil{Astronomy Department, Boston University, Boston, MA 02215}

\begin{abstract}

In a search of proper motion catalogs for common proper motion 
stars in the field of the
Kepler spacecraft I identified 93 likely binary systems. A 
comparison of their rotation periods 
is a test of the gyrochronology concept.
To find their periods I calculated the autocorrelation
function of the Kepler mission photometry for each star.
In most systems {\bf for which good periods can be found}, 
the cooler star has a longer period
than the hotter component in general agreement with models. {\bf However,} 
there is a wide
range in the gradients of lines connecting binary pairs in a period-color
diagram.
Furthermore, 
near the solar color, only a few stars have longer periods than 
the Sun, suggesting that they, and their cooler companions
are not much older than the Sun. In addition,
there is an apparent gap
at intermediate periods in the period
distribution of the late K and early M stars. Either {\bf star formation in this
direction has been variable, or}
stars evolve in period at a non-uniform rate, or 
some stars evolve more rapidly than others at the same mass.  Finally, 
using the autocorrelation function as a measure of the
activity level, I found that while the F, G and early K stars 
become less active as their periods increase,
there is no correlation between period and
activity for the mid K to early M stars.

\end{abstract}


\keywords{binaries: visual -- proper motions -- stars: activity -- 
stars: late-type -- stars: rotations -- stars: solar-type}

\section{Introduction}

It is by now well-established that stars with convective envelopes spin down
as they age. Since the much-cited paper by \citet{skumanich72} showing that 
solar-type stars follow an
approximate $t^{1/2}$\ relation for rotation period, a great deal of effort
has been made to understand the underlying physical mechanisms
and to derive a
quantitative calibration valid
for all cool, convective stars.  For some
recent discussions of stellar activity and
stellar rotations, see \citet{barnes10a}, \citet{barnes10b}  
and \citet{reinhold13}.  
The Sun and stars in young open clusters have been used as calibrators for the
age-period-color relation \citep[e.g.,][]{meibom15}, but it 
has been more difficult to find suitable
calibrating stars to probe the rotations of
older stars, particularly ones 
of later spectral types, \citep[but see][]{mamajek08, 
epstein14}.  

Efforts to provide a theoretical framework for stellar spin-down include 
\citet{kawaler88}, \citet{barnes07} and 
\citet{barnes10b}.  The key parameter is the convective turnover timescale
\citep{noyes84},
a quantity not directly observable, that drives the magnetic activity level,
which in turn controls the stellar wind that carries away angular momentum.
\citet{epstein14} have used a combination of models plus data from the open
cluster M37 and binary stars to show that differential rotation and (in stars
less than 0.6 $M_{\sun}$) the initial rotation rates
are the limiting 
factors in the age precision of older stars.  

The Kepler mission to search for exoplanets \citep{borucki11}
has for the first time provided the observational material needed
to extend the
calibation of the age-rotation period-spectral type relation to older and
cooler stars.  \citet{meibom11, meibom15} have used Kepler data to calculate
periods of a number of stars in the one Gyr cluster, NGC 6811,  
and the 2.5 Gyr cluster, NGC 6819. A 
much older cluster in the Kepler field, NGC 6791, is too distant to be 
useful for
rotational measurements of the main sequence stars, but in the extended
Kepler Mission (the ``K2 Mission'') the 4.0 Gyr cluster, M67, 
was observed,
and \citet{barnes16} have derived periods for some of the cluster stars. 
However, all of these clusters are relatively distant,
so cluster stars later than early K spectral type are too faint for Kepler.
\citet{angus15} used rotation periods of Kepler stars as well as
cluster stars and other stars with known ages to derive a period-age-color
relation of the form $P = a A^n (B - V - c)^b$, where
A is the age of a star in Myr and a, b, and n are constants.  They found
however that some stars exhibited large deviations from the mean relation.
 
\citet{mamajek08}, compared the
rotation periods of the components of a small sample of visual double stars;
while the actual ages of the binary systems may not be known, their 
components can
be expected to have the same age.  \citet{donahue98} had earlier 
applied the same idea
to compare the chromospheric ages of visual binary star components.
The Kepler mission has now made it possible to expand on the Mamajek and
Hillebrand and Donahue ideas and 
fill in the gaps in the 
age-color-period relation with the help of binary stars.  
In the Kepler field, there should be a 
considerable number of common proper motion (CPM) stars, stars moving through
space together; their components presumably have a common, if unknown, age.
\citet{deacon16} searched the Kepler field for wide binaries in
the Kepler field, to learn whether wide multiplicity can affect 
planet occurrence.  They found a number of likely binaries but only
briefly considered thw question of gyrochronology.
The aim of this project is to identify likely CPM stars in the Kepler field 
and use them to explore the existing age and spectral-type calibrations.

Section 2 is a description of the search process to find candidate CPM pairs 
and section 3 presents a confirmation of their binary nature from photometric
data. The procedure for finding rotation periods from
the Kepler photometry is the topic of section 4 and section 5 is a
discussion of the period distribution of proper motion pairs.

\section{Proper Motions}

There are at least 6 proper motion catalogs covering stars in the Kepler
field that might be of help to identify CPM stars in this region: 
The USNO-B catalog \citep{monet03},
the UCAC-4 catalog \citep{zach13}, 
the URAT-1 catalog \citep{zach15},
the Hipparcos/Tycho catalogs (specifically the Tycho-2 catalog) 
\citep{hog00},
the PPMXL catalog \citep{roeser10},
and the \citet{deacon16} study mentioned in the introduction that 
specifically targeted proper motions in the
Kepler field.  
Although these catalogs are not fully independent of one another, and have
a substantial overlap in the sources of positions from various epochs, 
there is nevertheless   
a considerable range in their 
completeness, magnitude coverage, astrometric and photometric precision and 
other characteristics.

To manage this
heterogeneous set of data, I downloaded the data from the first five of
the catalogs from the VizieR online database 
(vizier.u-strasbg.fr/viz-bin/VizieR).  
for an 8-degree radius field centered
at $RA = 19^h 22^m 40^s, Dec = 44^d 30' 00"$, which covers the entire Kepler
field of view. I also downloaded the complete \citet{deacon16} catalog which
covers approximately the same field.  

Together, these catalogs include several million stars, but only 206150 stars 
were observed by Kepler.  After downloading the 
positions,
physical properties and photometry from the MAST data
archive (archive.stsci.edu/kepler/), 
for all the stars observed by
Kepler, I merged these data together with a revised set of effective
temperatures and gravities for the Kepler list by \citet{huber14}.  I then
cross-referenced the merged MAST plus \citet{huber14} data with 
all six catalogs to get a complete 
list of proper motions, photometry and other information 
for the Kepler targets.  To match stars from any
two of the 7 lists, including the Kepler target list, the positions must match
within 2 arcseconds and the brightnesses must match within one magnitude.
Although they are on somewhat different photometric systems, each of the 
catalogs includes a red magnitude reasonably close to the SDSS r filter.

While the tabulated proper motion uncertainties in these catalogs are typically 
a few milliarcseconds
per year, the actual proper motion {\it differences} from 
one catalog to the other are often substantially larger.  For that reason,
I did not use the published proper motion uncertainties, except for those in the
Tycho catalog, in the following 
analysis unless the published uncertainty 
is larger than 10 msec/yr; in such cases,
I disregarded that proper motion value altogether.

Most of the stars with Kepler photometry match up with entries in more than 
one proper motion catalog; some selection process is necessary to create a
merged list of proper motions.  The Tycho proper motions are presumably the 
most reliable, so if there are Tycho values for a star, I chose to use them, 
including the published Tycho uncertainties. For the fainter
stars, I took the mean of the available proper motion values, giving double 
weight to
the Deacon catalog, because it is based on a combination of modern CCD 
photometry  and several earlier catalogs.  The Tycho 2 stars are all 
brighter than about r = 13 and the \citet{deacon16} catalog contains only
stars fainter than r = 14.5.
If there were three or more measurements and if one of
the values differs by more than 25 msec/yr from the mean of the others,
I tossed that value out.  
The resulting merged proper motions for the Kepler stars  
represent a compromise of the available
data; for estimates of the uncertainties 
I simply took the standard deviation of the
individual proper motion values.

The goal is to search for pairs of stars that appear to be 
moving together through space, that is to say, stars with a common proper
motion.  The typical proper motions of stars in this direction in the 
galaxy are close to zero, so for example, the 
mean proper motions of stars in the Tycho catalog are -0.4  
 and -3.1 milliarcseconds per year in RA and Dec respectively with 
dispersions of 9.7 and 14.8 milliarcsec per year.  This
means that for two stars to be reliably identified as likely CPM pairs their
proper motions must be substantial.  But stars with large proper motions
tend to be nearby, mostly low-luminosity dwarfs; in fact such 
stars are exactly the ones of interest for
this project.

Given a target star in some direction in the galaxy, the number of stars 
within an angular distance, $r_c$, of the target is 
$\onehalf \pi r_c^2 \rho(b)$, 
where $\rho(b)$ is the star density at galactic latitude $b$. (The factor $1/2$
accounts for the fact that for each target star, the search for companions 
proceeded only in the leading direction toward increasing RA - stars in the 
trailing direction would already have been searched.)  In general, almost
all stars are unrelated to the target star.  The probability
that a target star randomly has an {\it unrelated} companion with a proper 
motion, $\mu$\  within some range,
$\delta(\mu)$, of the target proper motion is $\sum{N(\mu)} / N_t$, where
$N(\mu)$\ is the density of stars in proper motion space at $\mu$ and $N_t$\ is
the total number of stars in the particular sample. The summation is
taken over the proper motion range $\pm \delta(\mu)$,  the proper motion
difference between the target star and its possible companion.

The total probability
of finding an unrelated star near a target star is just the product of the
factors:
\begin{equation}
\label{probable}
P(\mu) = \rho(b) \pi r_c^2 {\sum^{+\delta(\mu)}_{-\delta(\mu)}\left[N(\mu)
\right] \over 2 N_t} 
\end{equation}
So given a target star, the angular separation and  the proper motion 
difference between the two stars can be used in the above relation to get
the probability that this combination could occur by accident.  

Several additional constraints on possible companions are needed: 

1) I limited the search radius to the range, 
$6.0 \le r_c \le 600\ arcsec$. Although stars with a
small separation are more likely to be binaries than wider pairs, blended
images are a significant problem both for the measured proper motions
and for the Kepler photometry.

2) The total proper motion of each star must be greater than 5 
times the total proper motion uncertainty, which typically is of the order
of 3 mas yr$^{-1}$; 
for stars in the merged list with no calculated uncertainty, I assumed an
uncertainty twice that value.

3) For the proper motion search range, 
$\delta(\mu)$, around the target star, I used the lesser of  
the actual proper motion difference between the target and its 
possible companion or 5 mas yr$^{-1}$. 
 
4) The total proper motion difference between the two stars must be 
less than 10\% of the sum of their total proper motions.

5) In evaluating Equation \ref{probable}, 
I took the number of stars brighter than
the fainter of the two candidate stars as the value for $N_T$.

Using the constraints described above and the additional condition to
consider only pairs with a  
probability, $P(\mu) < 0.001$  that they could be an accidental pair, 
a search of the working proper motion list yields 
102 pairs of stars.  As a test of the selection criteria, I repeated the
search by successively adding and subtracting 20 arcminutes to the 
declination of each star before checking for matches to the target stars.
These tests yielded 19 pairs when the declinations were shifted northward and
13 stars with a southward shift.  Thus, as many as 
15\% of the pairs in the
original search could be accidental pairs, but as shown in the next section
photometric considerations lead to the elimination of additional stars.

Twelve of the candidate CPM stars are listed in the MAST archive as
``Kepler Objects of Interest'' (KOIs).  Two stars have confirmed planetary
systems - K06678383 (Kepler-104,b,c,d) and K12068975 (Kepler-197 b,c,d,e).
and three more stars are listed as having candidate planets - K08845205,
K09579208 and K11098013.  Six stars are listed as ``false positives'' -
K07013635, K07582691, K07885518, K07885570, K09762514, 
and K09762519. Three 
stars (K07885570, K09762519 and K06678383)
are also listed in the Villanova eclipsing binary catalog \citep{kirk16}. 

\section{Photometry}

The MAST data archive 
includes data for the Kepler stars on several photometric systems as well as
photometrically derived effective temperatures, but
many stars are missing values in one or more colors, in particular, the
B and V magnitudes.  However, the
Kepler index $g_{KIC}$,
which closely approximates the SDSS $g$\ index, and the 2-MASS $K_s$\ index 
are available in the archive 
for all but a few of the CPM stars.  Furthermore,
the color index, $(g_{KIC} - K_s)$, has good sensitivity to stellar temperature
over a wide range of spectral types and is widely available in various
stellar archives.  In the following, I will refer 
to these indices as ``$g$'' and ``$(g-K)$'' respectively.

\begin{figure}

\plotone{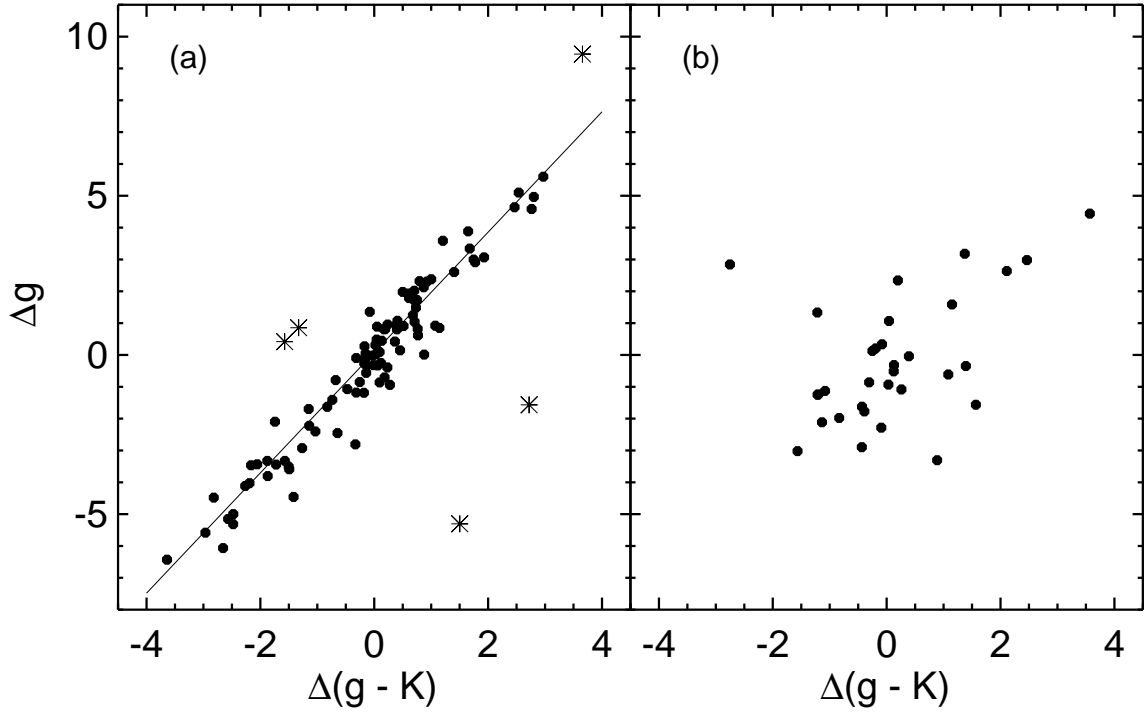}

\caption{A comparison of the magnitude differences, $\Delta g$, 
vs the color index differences, $\Delta (g-K)$. \label{cidiffs} 
(a) Candidate CPM pairs. 
(b) The two sets of false positives (created by adding or subtracting 20 
arcmin to each target star before searching for companions). The asterisks in 
Figure \ref{cidiffs}a represent stars with photometry incompatible with
being main-sequence common proper motion pairs. The regression line has
the form $\Delta g = 1.890 \Delta (g-K) + 0.079$.}

\end{figure}

If indeed both members of a suspected CPM pair are main sequence stars at
the same distance, then the 
magnitude difference between the two stars and their color index
difference should be strongly correlated. 
The correlation may not be a simple one:  
the main sequence is not linear;
some of the more massive stars may be evolved; some of the youngest, 
low-mass stars
may not have reached the main sequence; or one or the other of the two
stars may itself be multiple.  
Nevertheless, 
there is a well-defined 
correlation between the $(g-K)$ color differences and
differences in the $g$
magnitudes for most of the CPM candidates as shown in
Figure \ref{cidiffs}a.
The regression line in Figure \ref{cidiffs}a has
the form $\Delta g = 1.890 \Delta (g-K) + 0.079$.
In contrast, there is no correlation between color and magnitude differences
for stars in the two false 
positive tests (offsets North and South)
added together (Figure \ref{cidiffs}b).  

Three stars in the
initial sample do not have $g$\ or $K$\ magnitudes, and their photometric data
are incomplete in other respects as well.  Six more pairs in Figure 
\ref{cidiffs}a, including one outside the plotted area, are obvious outliers.
They are either not physically related, or one of the components may be 
an evolved star. Several others are early-type stars, outside the range of 
interest for this project.
I have rejected both components of 
these nine pairs from the CPM list.  For completeness, they are presented in
Table \ref{rejects}, which lists their KIC numbers, proper motions and 
uncertainties,
angular separation, g magnitude and $(g-K)$\ color index.

Several stars in the candidate list have Kepler 
gravities consistent with being evolved stars.  Indeed two stars with 
gravities typical of red giants
are among the outliers in Figure \ref{cidiffs}. 
Three other possibly low-gravity stars
have gravities characteristic of supergiants, but their proper motions,
their positions in Figure \ref{cidiffs} and, in two cases (K08888543 and
K08006740) their light curves, indicate that they are completely normal 
dwarfs. The third star of this group (K11503111) has an anomalous light
curve, suggesting some instrumental problem.  The \citet{huber14} gravities for
4 additional stars (K09468112, k08098181, K02992960, K02992956) indicate
that they may be slightly evolved, but they also fit near the fiducial line
in Figure \ref{cidiffs}.  The first of these four stars has properties
consistent with its partner, but the second star is a member of a close pair
with mutually compromised light curves. 
The last two of these form a binary pair with almost
identical properties, including their periods; they could be slightly evolved.
Since this group of 
stars listed 
with low gravities in \citet{huber14} fit the regression line in Figure
\ref{cidiffs} they remain in the list for now.  
The remaining stars have gravities consistent with their locations on the
main sequence.

The final list of candidate CPM
stars includes 93 pairs; they are presented 
in Table \ref{pairs}. The first column
in Table \ref{pairs} is a running pair number; column (2) represents the KIC
number from the Kepler input catalog; the next four columns represent the 
RA and Dec proper motions and their uncertainties, 
column (7) is the angular 
separation of the two components, column (8) is an index of the relative 
probability that it is a common proper motion pair on a scale where
a 99 indicates 
$P(\mu) = 0.00099$ (see Equation \ref{probable}), column (9)
is the g magnitude, and column (10) is
the $g-K$\ color. The remaining 
columns will be described in the following section.

\begin{deluxetable}{crrrrrrrr}
\tabletypesize{\scriptsize}
\tablecaption{Stars with Similar Motions Rejected as Common Proper Motion Pairs
\label{rejects}}
\tablewidth{0pt}
\tablehead{
\colhead{CPM pair} & \colhead{KIC} & \colhead{$\mu_\alpha$} & 
\colhead{$e_{\mu_\alpha}$} & \colhead{$\mu_\delta$} &
\colhead{$e_{\mu_\delta}$} & \colhead{Sep} & \colhead{$g$} & \colhead{$(g-K)$}
\\
\colhead{ } &
\colhead{ } &
\colhead{msec/yr} &
\colhead{msec/yr} &
\colhead{msec/yr} &
\colhead{msec/yr} &
\colhead{arcsec} &
\colhead{mag} &
\colhead{mag} 
}
\startdata
 11 &  7940959 & -28.4 & 0.8 &  -34.9 & 0.9 & 160.2 &  8.482 & 3.481 \\
 11 &  7941056 & -23.9 & 1.2 &  -32.2 & 1.9 & 160.2 & 13.787 & 2.200 \\
 28 &  9942231 &  -6.9 & 0.0 &   89.9 & 0.0 &  58.7 &  0.000 & 0.000 \\
 28 &  9942242 &  -5.8 & 1.3 &   86.1 & 1.3 &  58.7 & 17.597 & 4.460 \\
 30 & 11603064 &  -8.2 & 1.3 &  -17.4 & 1.3 &  94.2 &  8.437 & 2.595 \\
 30 & 11603098 &  -7.6 & 1.5 &  -20.6 & 1.5 &  94.2 & 10.003 & 0.513 \\
 41 & 12645102 &  -5.8 & 2.5 &  -21.8 & 2.2 &  54.7 & 12.806 & 1.309 \\
 41 & 12645107 &  -7.2 & 2.7 &  -18.6 & 2.5 &  54.7 & 11.951 & 2.303 \\
 44 & 12254819 &  21.4 & 0.0 &   62.6 & 0.0 &   7.9 &  0.000 & 0.000 \\
 44 & 12254821 &  20.4 & 0.7 &   68.5 & 4.5 &   7.9 & 13.116 & 3.345 \\
 48 &  7816381 &  -6.2 & 1.6 &   45.8 & 1.7 &  14.3 & 17.688 & 3.212 \\
 48 &  7816387 &  -8.7 & 1.3 &   47.4 & 1.9 &  14.3 & 13.153 & 1.449 \\
 56 &  8557567 & -76.5 & 0.0 & -132.8 & 0.0 & 302.5 & 18.512 & 4.799 \\
 56 &  8557784 & -71.3 & 1.0 & -130.0 & 1.0 & 302.5 &  9.062 & 1.628 \\
 75 & 11822514 &  25.0 & 1.3 &   44.6 & 1.7 &  29.4 & 17.528 & 4.926 \\
 75 & 11822535 &  23.0 & 1.3 &   46.4 & 1.0 &  29.4 & 14.455 & -0.913 \\
 83 & 11774619 & -28.0 & 0.8 &  -31.7 & 1.6 &  37.6 & 16.567 & 2.780 \\
 83 & 11774639 & -24.5 & 1.0 &  -26.8 & 0.8 &  37.6 & 16.148 & 4.758 \\
\enddata
\end{deluxetable}

\begin{deluxetable}{crrrrrrrrrrrrrrrc}
\tabletypesize{\scriptsize}
\rotate
\tablecaption{Selected Common Proper Motion Stars in the Kepler Field
\label{pairs}}
\tablewidth{0pt}
\tablehead{
\colhead{CPM pair} & \colhead{KIC} & \colhead{$\mu_\alpha$} & 
\colhead{$e_{\mu_\alpha}$} & \colhead{$\mu_\delta$} &
\colhead{$e_{\mu_\delta}$} & \colhead{Sep} & \colhead{I$_{RP}$} & 
\colhead{$g$} &
\colhead{$g-K$} & 
\colhead{Period} & \colhead{e$_{Period}$} & 
\colhead{V} &
\colhead{$\sigma$V} & 
\colhead{N$_{qtrs}$} & 
\colhead{Type} \\
\colhead{ } &
\colhead{ } &
\colhead{msec/yr} &
\colhead{msec/yr} &
\colhead{msec/yr} &
\colhead{msec/yr} &
\colhead{arcsec} &
\colhead{ } &
\colhead{mag} &
\colhead{mag} &
\colhead{days} &
\colhead{days} &
\colhead{ppm } &
\colhead{ppm } &
\colhead{ } &
\colhead{ } & \\
\colhead{(1)} &
\colhead{(2)} &
\colhead{(3)} &
\colhead{(4)} &
\colhead{(5)} &
\colhead{(6)} &
\colhead{(7)} &
\colhead{(8)} &
\colhead{(9)} &
\colhead{(10)} &
\colhead{(11)} &
\colhead{(12)} &
\colhead{(13)} &
\colhead{(14)} &
\colhead{(15)} &
\colhead{(16)} &
}
\startdata

  1 &  7090649 &  16.1 & 0.6 &  10.8 & 2.5 & 9.3 & 99 & 11.976 & 2.277  
& 8.27 & 0.11 & 6705.7 & 5.52 & 17 & P & \\
  1 &  7090654 &  17.6 & 1.1 &  13.8 & 1.1 & 9.3 & 99 & 10.249 & 1.528 
& 1.98 & 0.00 & 2578.3 & 52.0 & 17 & P & \\
  2 &  7659402 &  28.5 & 0.5 &   8.9 & 1.1 & 29.0 & 56 & 16.708 & 4.593
& 19.66 & 0.28 & 801.1 & 8.2 & 4 & A & \\
  2 &  7659417 &  27.2 & 0.9 &   7.9 & 1.5 & 29.0 & 56 & 17.559 & 4.848
& 16.31 & 0.53 & 44.8 & 5.0 & 17 & P & \\
  3 &  7582687 & -26.5 & 1.5 & -25.0 & 1.0 & 21.8 & 93 & 13.066 & 2.172
& 14.07 & 1.94 & 46.1 & 2.4 & 17 & P & \\
  3 &  7582691 & -23.4 & 3.0 & -23.8 & 0.6 & 21.8 & 93 & 16.499 & 4.227
& 24.54 & 0.12 & 642.7 & 5.8 & 14 & A & \\

\enddata

\tablecomments{Table \ref{pairs} is published in its entirety in the 
electronic edition of the {\it Astronomical Journal}.  A portion is 
shown here for guidance regarding its form and content.}

\end{deluxetable}

Because most of the stars that satisfy the present criteria are relatively
nearby, they should be expected to have low interstellar reddening.  In
any case, since the two components presumably lie at the same distance,
the reddening would be expected to be the same, and so will not affect 
Figure \ref{cidiffs}.  

The
Kepler data archive does include reddening values for many of 
the Kepler targets, so to 
correct the photometry for the effects of reddening, I corrected each pair
by the average reddening of the two stars..  In cases where one (or both) of 
the  members did not have a reddening value, I assigned a value of $E(B-V) = 
0.06$ (the mean value of the stars in this sample with reddening values), 
before calculating the average value for that pair.  
To correct the 
$(g-K)$\ and $g$\ indices, I applied the extinction coefficients described by 
\citet{yuan13}.  With help of their Table 2, I found that 
$A(g) = 3.30 E(B-V)$ and 
$E(g-K) = 2.99 E(B-V)$.

\section{Rotation Periods}

The Kepler PDC-MAP long-cadence data for each star (obtained from MAST)
consists of a nearly continuous series of brightness measures
with a cadence of about 30 minutes ($\sim$0.0204 days), covering a period of
4 years. The data for each approximately 90-day period, or ``quarter,''
were processed 
separately, so there are differences in photometric zero-points and other 
parameters from quarter to quarter.  
As described in the archive, each quarter's data have
been independently processed to remove most instrumental artifacts, while 
hopefully having only a minimal effect on the astrophysical phenomena.  

Assuming that the ubiquitous brightness variations of cool main sequence stars 
are the result of rotational migration of starspots across the visible 
hemisphere, it should be possible to identify the rotational period with 
standard time series analysis methods.  Following \citet{mcquillan14} I 
computed the autocorrelation function (ACF) of each star's PDC-MAP lightcurve.

\begin{figure}

\plotone{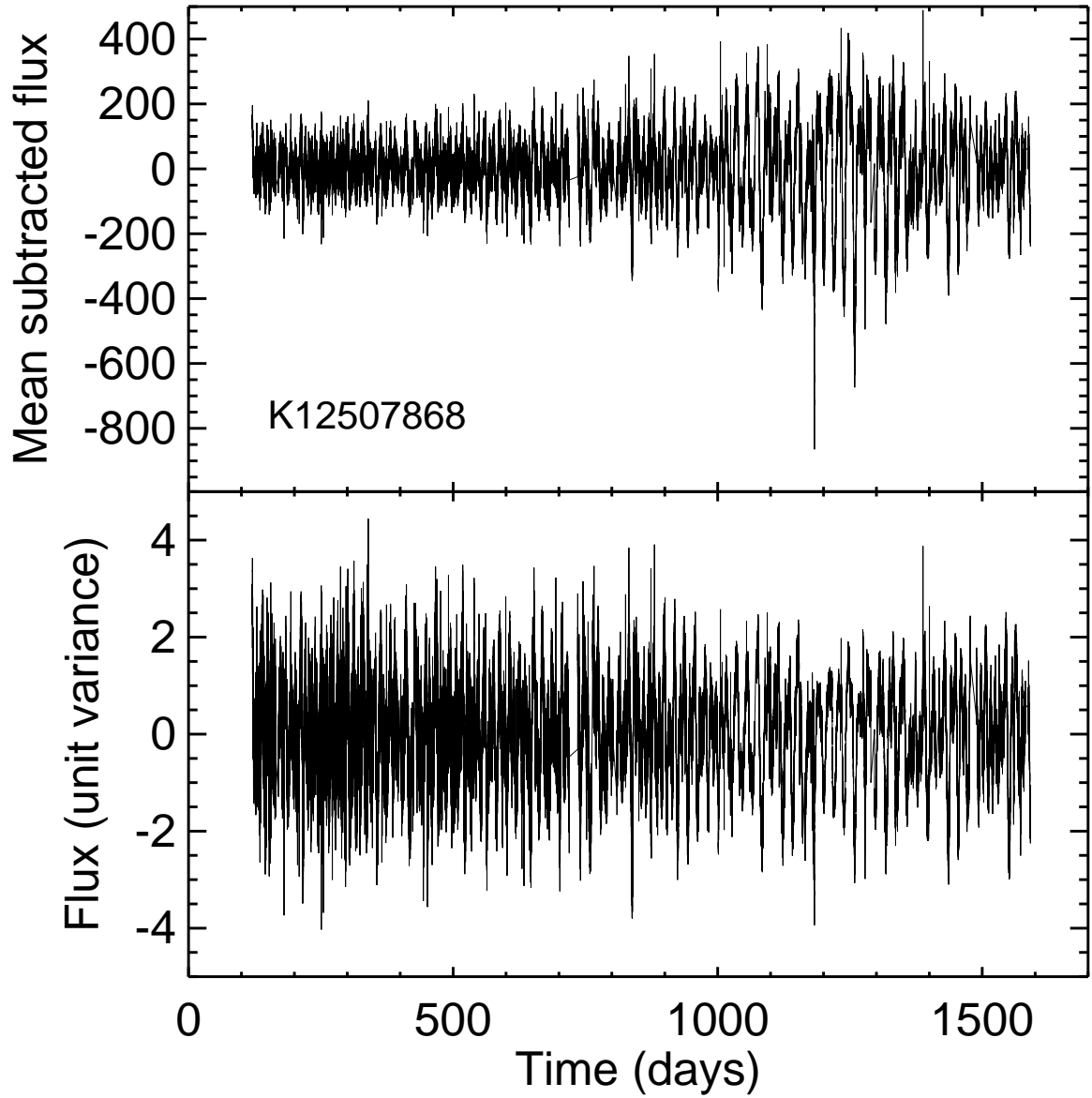}

\caption{Star K12507868.  The upper figure shows the 17 quarters of 
PDC-MAP data with the
quarterly mean values subtracted.  In the lower figure, the quarterly 
data values have also been
normalized to unit variance by dividing by the quarterly RMS value.
\label{rescaled}}

\end{figure}

To compute the ACF, I began by subtracting 
the mean value of each quarter's measures and dividing by the quarterly rms
value.
\begin{equation}
\label{pdcmean}
P(i,j) = \left[F(i,j) - \bar{F}(j)\right] / \sigma_{\bar{F}(j)},    
\end{equation}
where $F(i,j)$\ is the measured flux of 
observation $i$ in quarter $j$, $\bar{F}(j)$\ is the mean value of the data
for quarter j and $\sigma_{F(j)}$\ is the rms value of the zero mean values
in quarter $j$.  So $P(i,j)$\ has a mean value of zero and unit variance 
separately in each
quarter.  But the variance of the original 
data, $F(i,j)$, will in general differ from quarter
to quarter, as
a result of both instrumental and astrophysical differences in the quarterly
data sets.  In particular, because of the small scale difference from quarter
to quarter and because of differences in the variance there will be (generally)
slight scale differences in the distribution of
measurement uncertainties 
from one quarter to the next. Figure \ref{rescaled} shows
one of the more noticeable cases of this phenomenon.  The upper diagram shows
the 17 quarters of data with the quarterly mean values subtracted; in the
lower figure the data have also been divided by the quarterly rms values.
The assumption here is that
these inconsistencies will be immaterial to finding periods.

The ACF, A(k), can be defined as
\begin{equation}
\label{acfeqn}
A(k) = {1 \over \sum_j\left[n_{jk} - k\right]} \sum_j \left\{\sum_i\left[P(i,j)
P(i+k,j)\right]\right\},
\end{equation}
where $k$\ is called the {\it lag}, measured as multiples of the 0.0204 day
observing cadence.
The number of useful
measures in each quarter at lag, k, is $n_{jk}$.  
Any periodic variations in the 
measured brightness of the star (possibly the rotation period) will 
generate an ACF with a series of peaks with lag values at multiples of the
period.  As Equation \ref{acfeqn} 
shows, the final value of $A(k)$ is the sum of the correlations
over all data segments and can vary between plus and
minus one.

By calculating the ACF in this way, three problems are overcome.  First, the
Kepler quarterly datasets are somewhat independent of one another, with 
slightly different scale factors and independent calibations. 
Second there are gaps of various lengths in the data; $n_{jk}$\ represents the
actual number of measures for a particular star in quarter j. 
Finally, while
the rotation periods are completely regular, the spot features, on which
the measurements  depend, change on time scales that may be shorter than
a single rotation, and are rarely completely correlated over many rotations.
So it is appropriate simply to add up a series of relatively short segments
that each are more-or-less
correlated over one or a few rotation periods.

\begin{figure}

\plotone{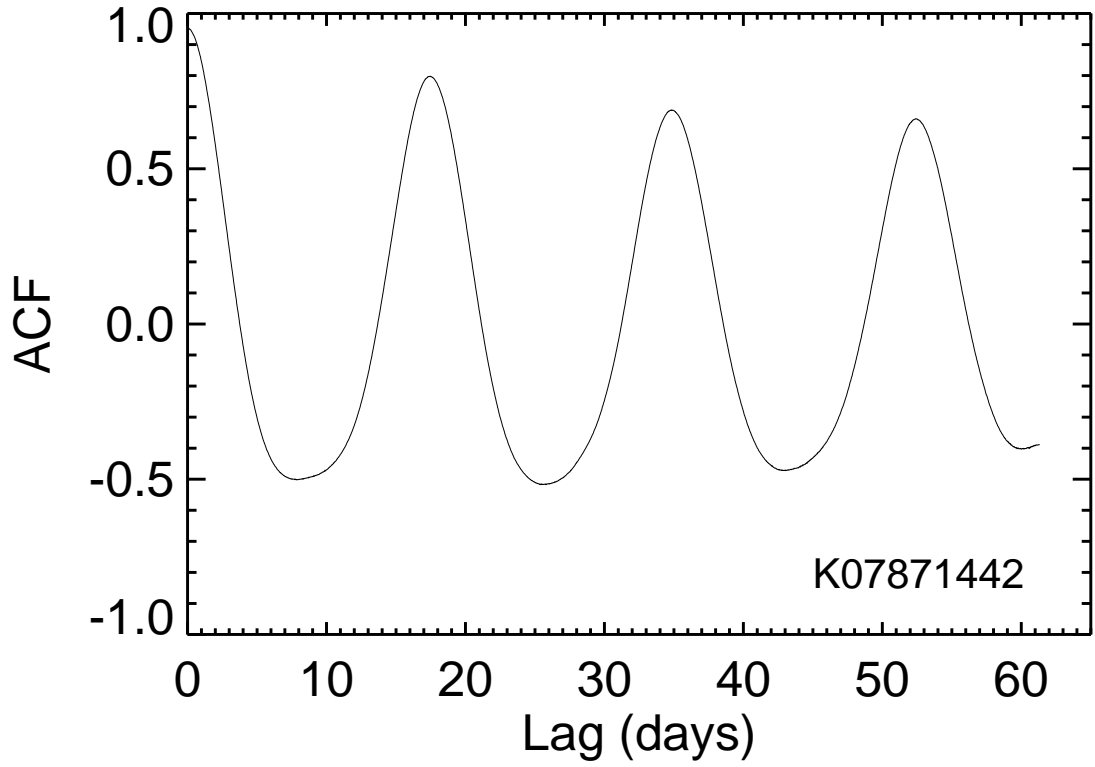}

\caption{Star K07871442, an early M star with a well-defined ACF, indicating 
a period of 17.45 days.  This curve covers 17 quarters of Kepler data.
\label{example1}}

\end{figure}

\begin{figure}

\plotone{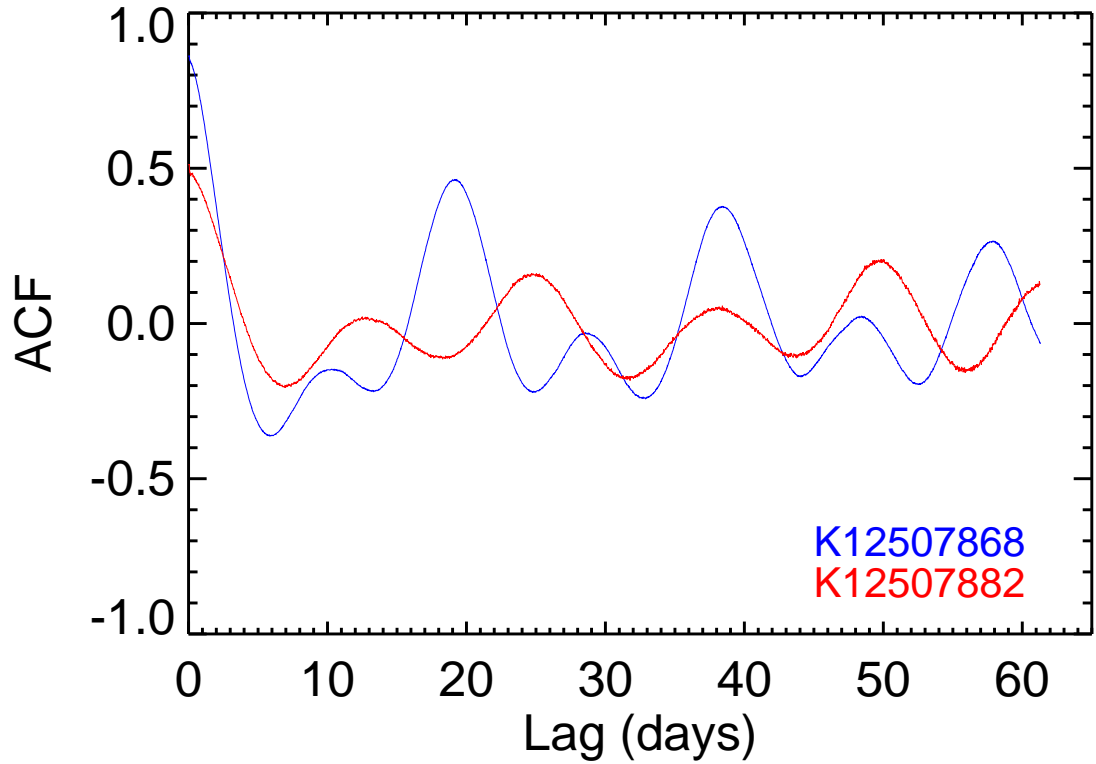}

\caption{Two components of a CPM pair, K12507868 and K12507882, both
of which have alternating peak heights.  
The periods derived here, 19.25 days and
24.82 days are in good agreement with the values found by \citet{mcquillan14}.
\label{example2}}

\end{figure}

\begin{figure}

\plotone{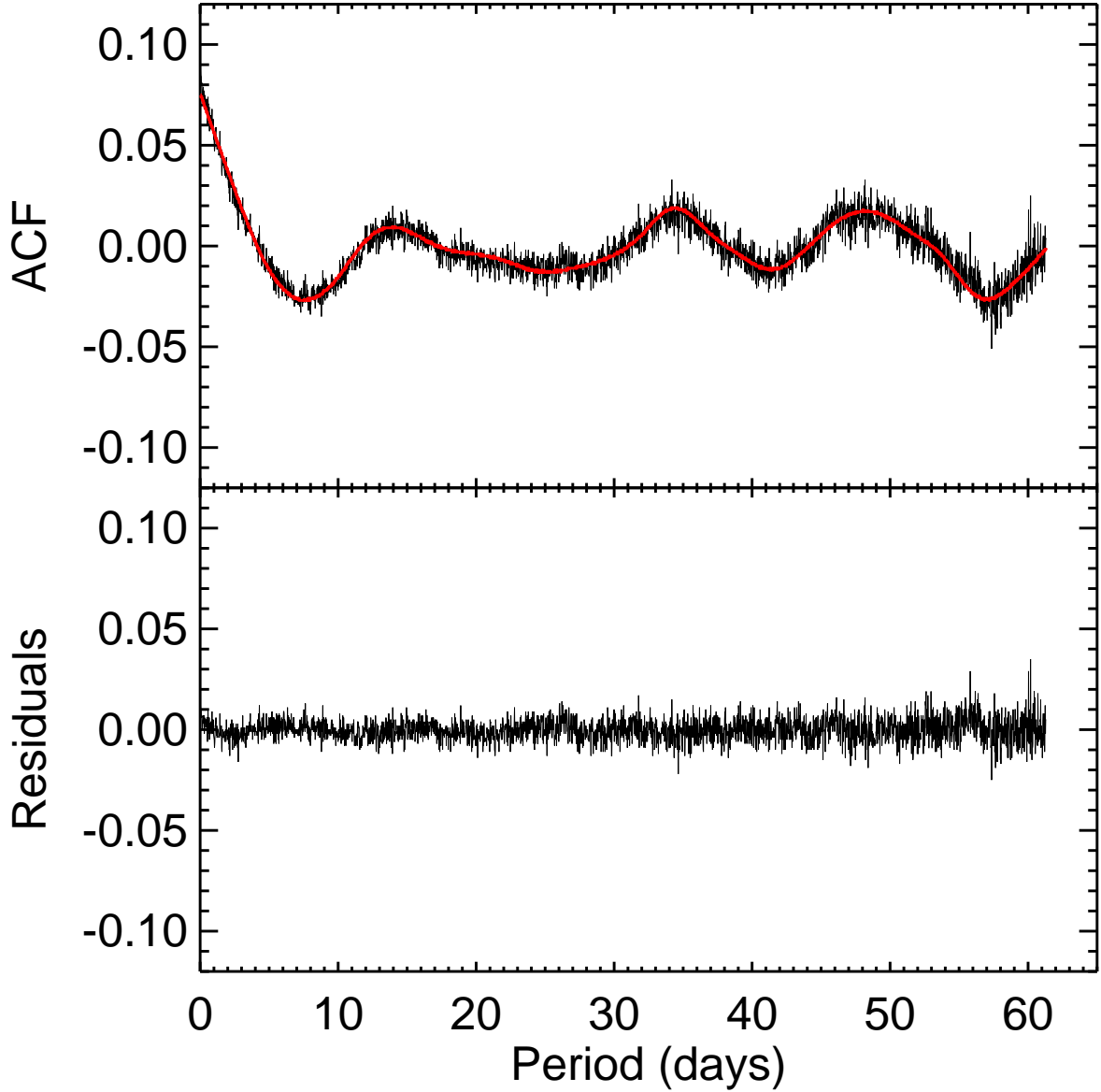}

\caption{Star K09762514, which has a rather noisy, low-amplitude ACF, showing 
limited evidence for activity.  The
red curve in the upper figure
is an illustration of the spline fitting 
to the ACF, with knot points at 2.5 day 
intervals.  The best period found here is 16.08 days. \citet{mcquillan14}
considered the period to be indeterminate, quoting a tentative 
value of 45.209 days. The lower curve shows the residuals between the
data and the spline fit. The peak amplitude of the ACF is 9.3 
$\pm$\ 2.8 (ppm).
\label{splinefit}}

\end{figure}

\begin{figure}

\plotone{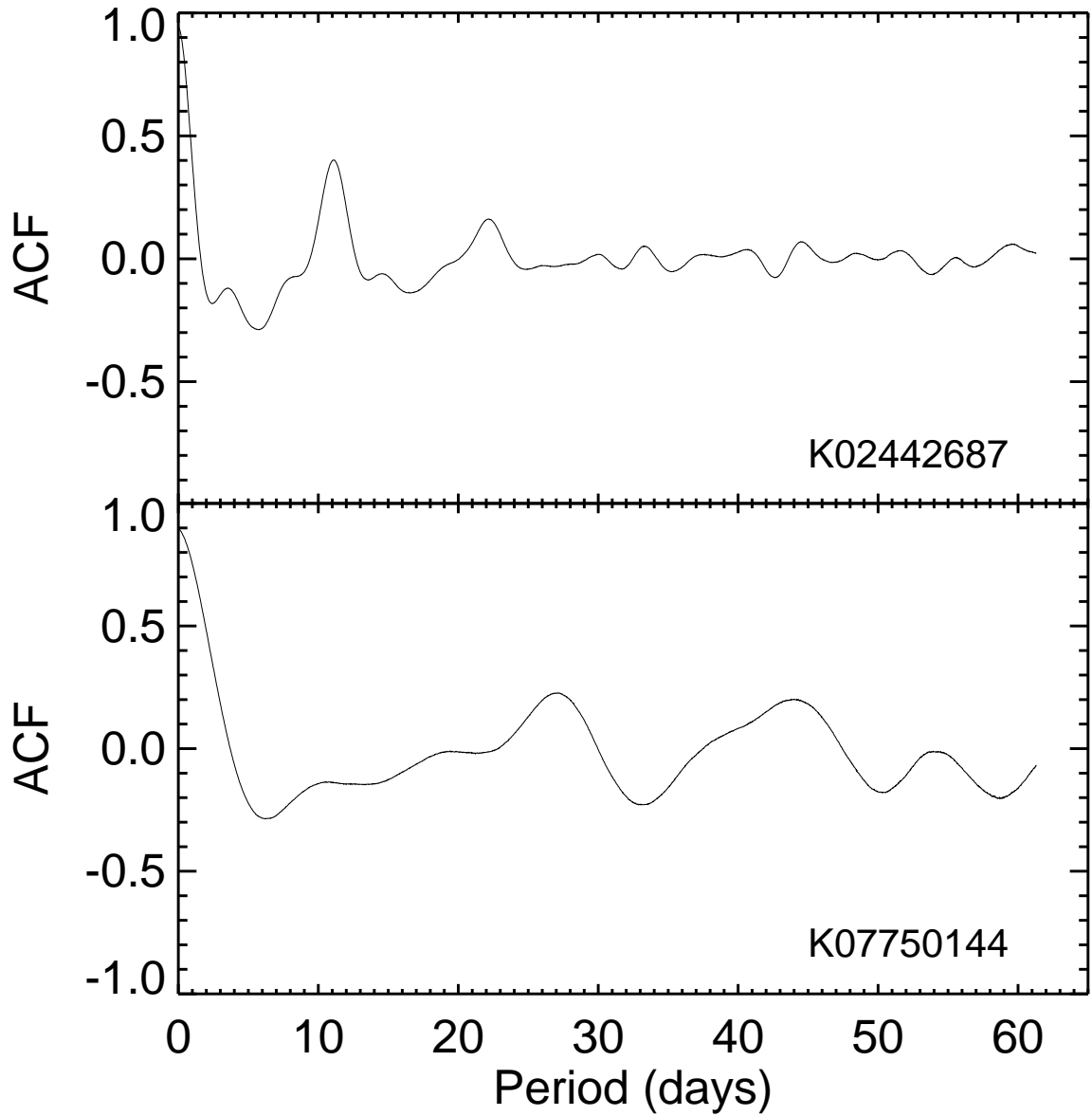}

\caption{Two stars with somewhat complex ACF curves.  K02442687 is an F star
with a possible period of 10.98 days; its ACF curve is typical for the 
earlier spectral-type stars.
K07750144 is a solar-type star with
a possible period of 27.08 days. 
The former star is outside the temperature
range searched by \citet{mcquillan14}, but they found a
period of 26.431 days for the latter. \label{example4}}

\end{figure}

Figures \ref{example1} - \ref{example4} illustrate the range of shapes 
taken by the ACF curves.  Active stars with long-lived surface features
exhibit regular patterns (Figure \ref{example1}), athough it is not always 
certain what the true rotation period is (e.g.,  Figure \ref{example2}).
Less active stars may show hints of periodic variation (Figure 
\ref{splinefit}),
although the variations could instead be primarily instrumental in origin,
particularly at large periods.  
Finally,
some stars have more complex behaviour (Figure \ref{example4}), 
and as mentioned 
above, several several stars are KOI's selected as possible planet hosts or
eclipsing binaries.

\begin{figure}

\plotone{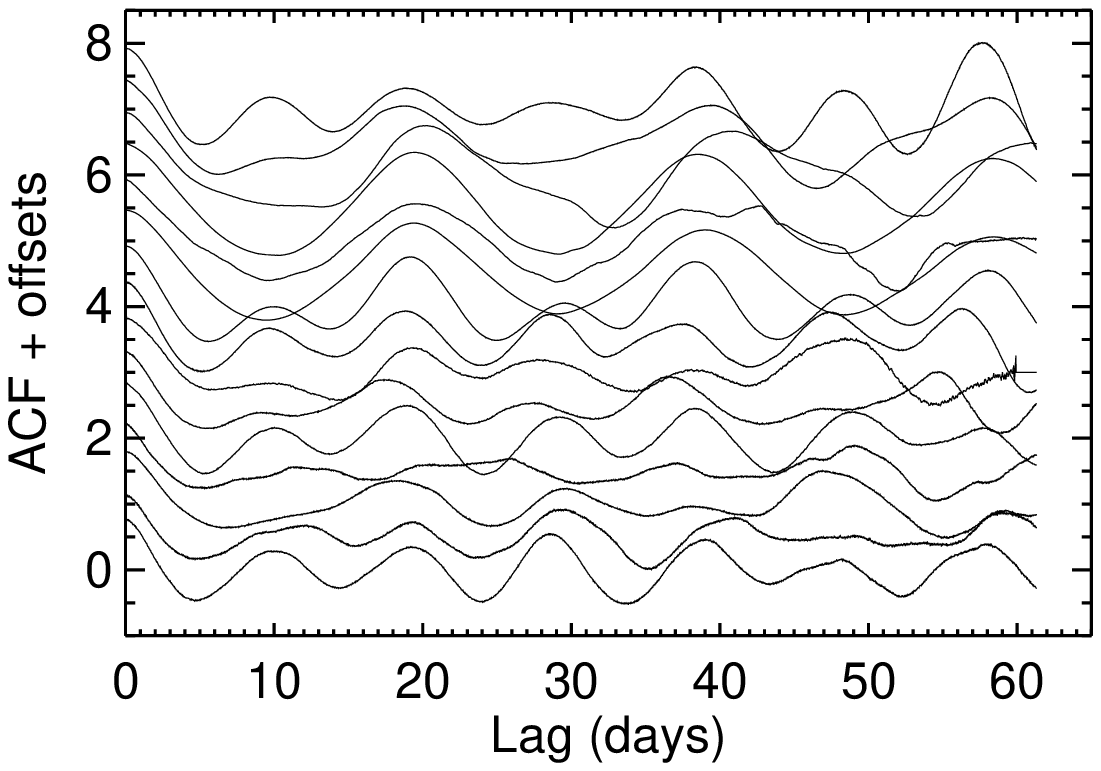}

\caption{ACF curves quarter-by-quarter for star K12507868. Successive 
quarters are offset by factors of 0.5 for display purposes.  The composite
curve over all quarters is shown in Figure \ref{example2} 
\label{quarterly}}

\end{figure}

Notwithstanding the smooth, repetitive patterns of Figures \ref{example1} and
\ref{example2}, There are in many cases large differences in the patterns from
quarter to quarter.  For example, Figure \ref{quarterly}
illustrates the
individual quarterly ACF patterns for star K12507868 (see Figure 
\ref{example2}).
There are considerable differences in the character of the variability from
quarter to quarter; it is only in the summation of $A(k)$ 
over several quarters that the dominant pattern (which 
presumably is the actual rotation period) becomes apparent.
For that reason, the rotation periods of stars with measures over only a few
quarters are often considerably less certain.

After calculating the ACF of a star, the next step is to find 
its period, or periods, by searching for peaks in the ACF.  First I
smoothed the ACF 
function by finding the maximum-likelihood cubic spline interpolation 
coefficents to fit a smooth curve through the ACF points.  In Equation
\ref{pdcmean} 
the individual data points, $P(i,j)$, are defined by 
the original measures, with a
cadence of approximately 30 minutes.  I chose the spline knot points
at multiples of 24 datapoints i.e. at multiples of about one-half day.  
Given a starting 
set of knot points, I used a simple Monte-Carlo process to find 
values for the spline coefficients that minimize the standard deviation between
the actual data points and the calculated spline function. Figure 
\ref{splinefit} illustrates the fitting of a  spline curve to a somewhat noisy,
low-amplitude ACF.

The fitting program chose the peaks, i.e., the local maxima in the smoothed 
ACF, and
the initial estimate of the rotation period was taken to be the mean of the
separations between the identifiable peaks in the smoothed ACF.
In some cases, after inspecting the individual ACF functions, I selected twice 
the initial period if there were
alternating peak heights (see, e.g., Figure \ref{example2}).
There are also several stars 
with a pattern of
peak heights repeating in groups of three with the third peak being
the highest. For the period uncertainty, I took the standard deviation of the 
period measures. 

The rotation periods and uncertainties are listed in columns 11 and 12 of Table
\ref{pairs} and the number of Kepler quarters used in the analysis are given
in column 15.  Stars with period uncertainties listed as 0.00 are all
short period stars with uncertainties less than 0.01 day. 

\begin{figure}

\plotone{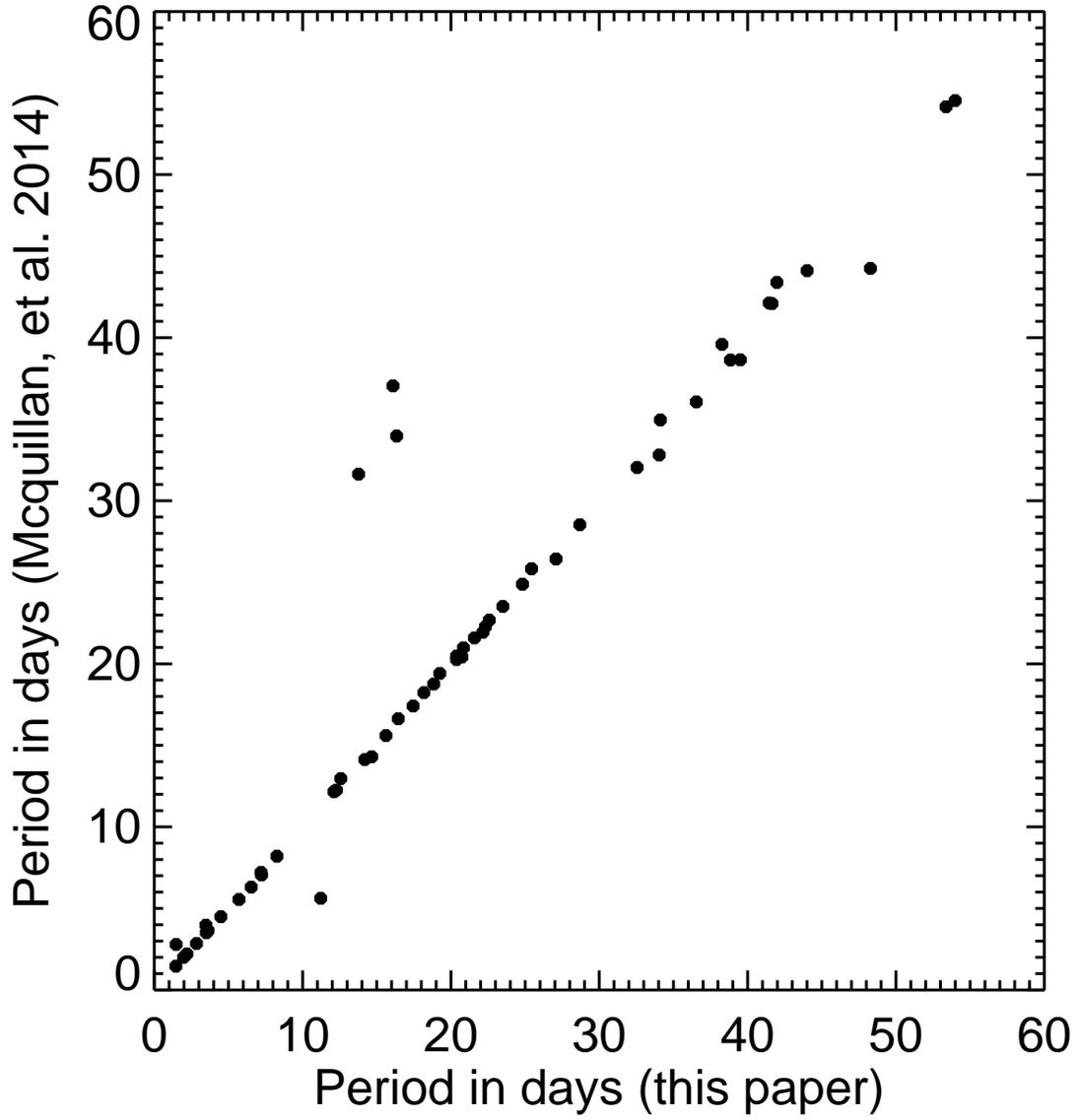}

\caption{Comparison between the rotation periods in this paper
and those of
\citet{mcquillan14}.
\label{pdiffs}}

\end{figure}

\citet{mcquillan14} did an autocorrelation analysis similar to the
present one.  After
searching 133,030 main sequence Kepler stars for periods, they were able 
to calculate rotation periods for 34,030 stars. 
Their full sample
included 120 of the 186 CPM stars in Table \ref{pairs}, but only 58 stars are
in common between Table \ref{pairs}  and the 34,030 stars for which they 
found good periods.  
In addition to excluding known red giants, they rejected 
KOI stars, ones with no effective temperature or gravity measures, stars
with $T_{eff} > 6500K$, and stars that were followed for fewer than 8
quarters.  
For the 58 stars in common
to the two studies, the agreement is generally very good, even though there
are some small differences in the procedures.  Figure \ref{pdiffs} shows the
differences between the McQuillan, et al. values and the values in Table
\ref{pairs}.  There are eight stars for which their periods disagree by 
more than a little over one day.  Star K10388259 has a very low amplitude
and long period,
so a difference of 4 days out of 48 is not surprising.  For three more stars
(K04544938, K04946401 and K10230145), the McQuillan, et al. periods are
twice the values in Table \ref{pairs} and for another star, the Table 
\ref{pairs} value is
twice the McQuillan, et al. period; either period is possible as the
actual rotation period for these stars.  
The remaining three stars (K04931390, K10388283,
and K11876220) all have two distinct, very regular, periods.  I find 
0.3 \& 7.62 days, 2.74 \& 38.82 days and 1.47 \& 20.83 days respectively for
these three stars.  They are not included in the figure.
For the remaining 50 stars, the difference is
-0.05 $\pm$\ 0.46 days.

\subsection{Stellar Activity}

The autocorrelation function (Equation \ref{acfeqn}) is normalized by the
variance of the data, so by definition, the ACF at zero lag has a value of
one.  In the case of a function consisting purely of random noise,
all other values of $A(k)$ would have a value of zero, within the noise level.
But if a function has at least some degree of correlation at 
periods greater than the sampling period, $A(k)$ will vary with $k$\ 
depending on the functional form and the degree of the correlation.
So the autocorrelation coefficient
simply shows the degree to which the data are correlated at some particular
value of $k$; when there are
peaks in the ACF, they indicate the presence of a periodic phenomenon, which 
could be either instrumental or intrinsic to the star.

Some commonly used measures of the activity level of stars in the 
Kepler program are derived from the variance of the data, such as the range
of variation between the 5th and 95 percentiles of variation
\citep[see][for example]{basri13}.
But this definition does not distinguish between random noise and
intrinsic variability.
With the help of the ACF, it is possible separate the variance into its
random and periodic fractions.  The ACF value at a peak represents
the correlated fraction of the variance at that period - so the product of
the total variance and the ACF value at the peak can be used as a measure of
the actual stellar variability:
\begin{equation}
\label{bigv}
V = 1000000 {A_{max} \over n_q} \sum_j\sigma_{\bar{F}(j)},    
\end{equation}
where $A_{max}$\ is the value of the ACF at the highest peak, $n_q$\ is the 
number of quarters this star was observed and
$\sigma_{\bar{F}(j)}$ is the rms value of the zero mean
of the data in quarter $j$, including both random noise and any 
intrinsic or instrumental signals as defined in Equation \ref{pdcmean}.
The result is a fractional periodic
amplitude in units of parts-per-million. 
For brighter stars $A_{max}$ may be close to 1, that is the quantity V is
dominated by the intrinsic variance of the data, and the purely random noise 
component
is small.   For stars that are almost constant, or for faint stars, $A_{max}$ 
may be (sometimes much) less than one so V is less than the total 
variance of the 
data.  To the
extent that correlated instrumental (non-astrophysical) 
signals can be ignored, V is a measure of the intrinsic variations of the star.
The uncertainty in V is just random component of the variance of the 
data; it is eqal to the standard deviation of the residuals of the 
cubic spline fit to the ACF curve (see, e.g. Figure \ref{splinefit}).
The variability measures, V, and $\sigma_V$\ are listed in columns 14 
and 15 of Table
\ref{pairs}.

Column 16
of Table \ref{pairs} describes the visual appearance of the ACF.
Stars with regular nearly equal peaks are labelled with a ``P,'' (as in 
Figure \ref{example1}), those with
alternating peak heights (see Figure \ref{example2}) are designated with an 
``A'' and triple-peaked stars with a ``T.''
The 31 stars with $V / \sigma_V < 5$\ are designated as ``I'' 
(indeterminate) in Table 
\ref{pairs}.  Those with indeterminate periods are either 
nearly constant or very faint or they were observed for only a few quarters.
All but one of the indeterminate stars with 
$(g-K) < 3.5$\ were observed for more than 10 quarters, but no reliable periods
could be found; they are simply inactive.  Nearly all of the cooler 
stars were observed for four
or fewer quarters, and although some of them showed signs of variability,
it was not possible to determine reliable periods.
Another 14 stars (designated with a ``C'' in Table \ref{pairs}) 
exhibited various pecularities in their ACF curves; several of
them have two distinct periods.  Most of the stars in this group were close 
pairs
with separations less than about 15 arcseconds and several of them 
interfere photometrically with one another. The C-type stars are not included
in the analysis.
The remaining 141 active stars, including 56 complete pairs, 
make up the primary sample for this analysis.  They are variable presumably
as a result of active regions moving across their surfaces as they rotate.

\begin{figure}
\plotone{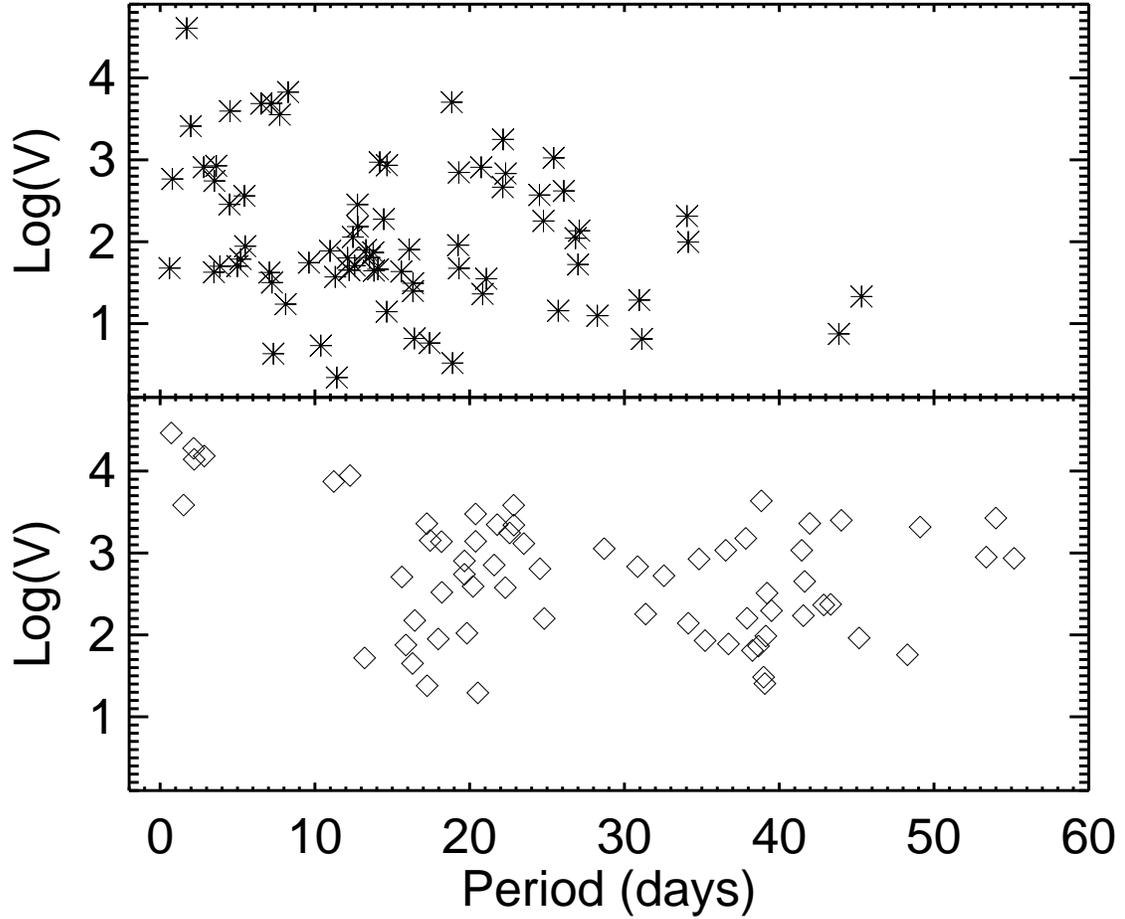}
\caption{\label{active}
The rotation periods vs activity levels 
of all active stars in the CPM sample are shown.  The asterisks (upper figure)
represent stars
with $(g-K) < 3.0$\ and the diamonds (lower figure) 
are stars with $(g-K) > 3.0$.  Stars labvelled ``I'' or ``C'' are not shown.}

\end{figure}

Figure \ref{active} shows the relation between rotation period and the log of
the activity index, $V$\ for all of the active stars.  The upper diagram
includes stars with $(g-K) < 3.0$\ ($T_{eff} \sim 4750K$, spectral type K3)
and the lower diagram shows cooler stars with 
$(g-K) > 3.0$.  
There is an expectation that as stars age, the level of activity decreases
and the rotation period increases (see the Introduction).
But as this figure shows, 
at every period there is more than a two order-of-magnitude
variation in the activity level.  The bluer stars do show a 
general declining 
trend in
activity with increasing period in agreement with previous work. However,
among the redder stars, except for a few of the shortest period stars which 
tend to have higher activity levels,
there is no correlation between activity and 
rotation period between about 10 days and 50 days.  Overall, the cooler stars 
have higher activity levels than the hotter stars at every period.

\section{Gyrochronology}

The relationship between stellar ages, rotation periods and colors (or
mass) 
for F, G, K, and M stars younger than about 600 Myr has been thoroughly 
studied using open clusters up to 
the age of the Hyades and Praesepe.  
Now, as described in the introduction, the age-period-color 
relation for F, G and early K
stars has been extended to the Sun's age, with the help of
open clusters observed by the Kepler mission.  But because the clusters are
all distant, the age-period-color relation for the cooler stars
(beyond early K spectral type) is still mostly unknown.  The 
present sample of common proper motion stars span
the range of late-type main sequence stars from mid-F spectral type to early
M stars.

\begin{figure}
\plotone{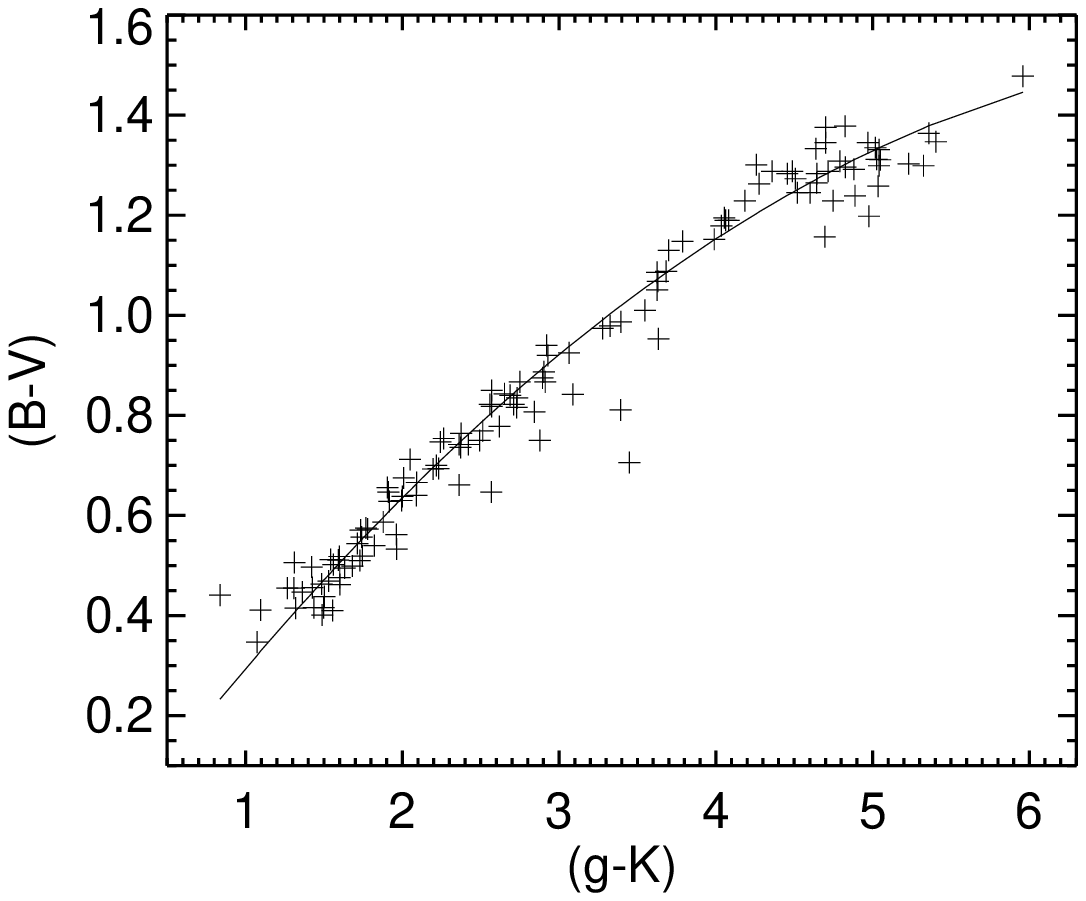}
\caption{The (g-K) vs (B-V) color indices from the MAST archive for all
stars in the present sample with tabulated colors.  The photometry has been
corrected for reddening using data from the archive.  For stars without 
reddening values, I used $E(B-V) = 0.06$. The solid line is a
simple quadratic least-squares fit of (g-K) to (B-V) after correcting for
reddening (Equation \ref{gkbv}).}
\label{gk2bv}
\end{figure}

\begin{figure}

\plotone{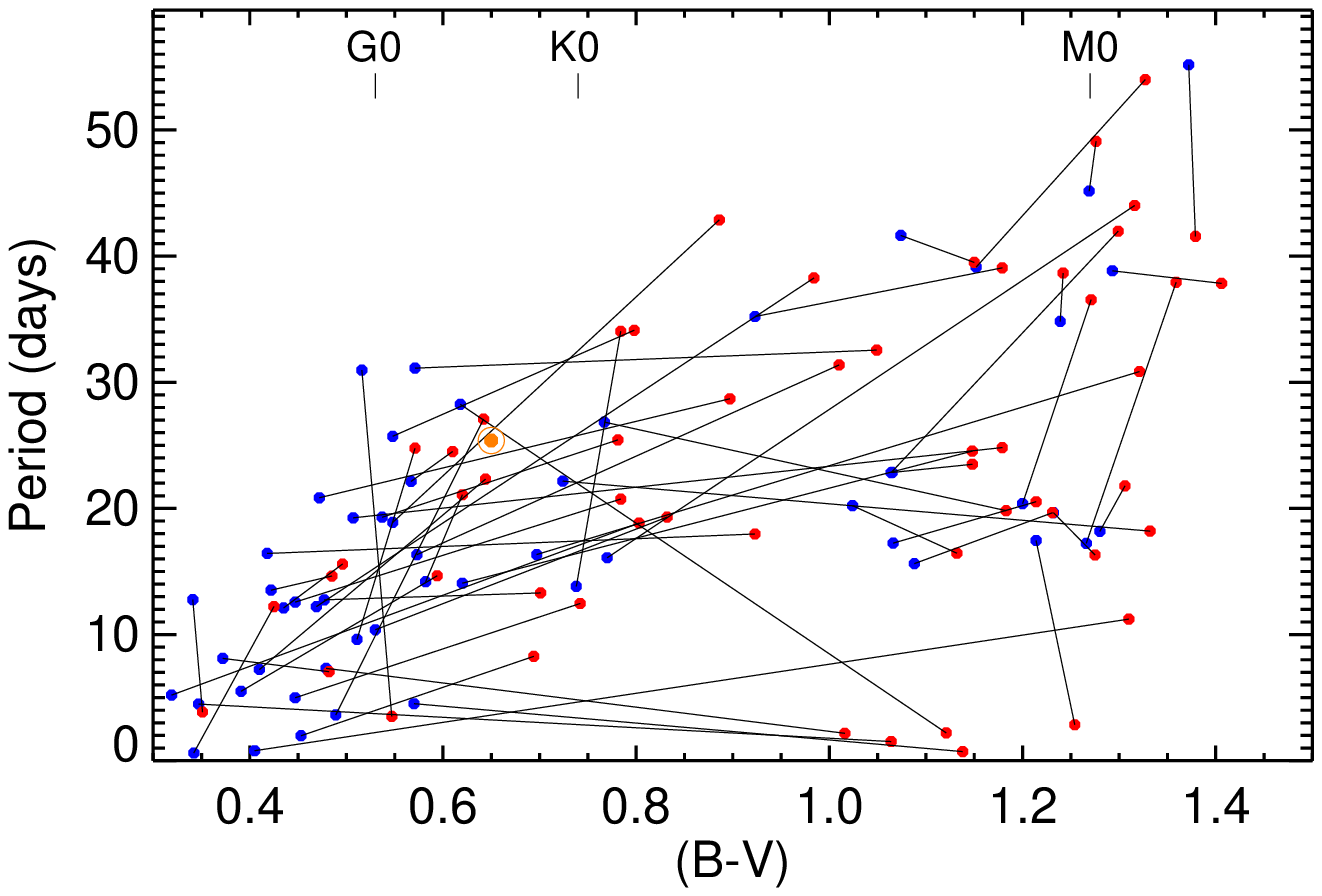}

\caption{Period-color diagram for CPM pairs with two active components. 
The lines connect the hotter star of each pair (blue symbols) with the cooler 
star (red symbol).  The (B-V) values are derived from Equation \ref{gkbv}.
The Sun's position is shown
with a dotted circle. The approximate locations of selected spectral types
from Table 5 of \citet{kraus07} are also shown. {\bf The blue component of one
active pair (number 80) lies outside the diagram so that pair is not shown.}
\label{gyro1}}
\end{figure}

\begin{figure}

\plotone{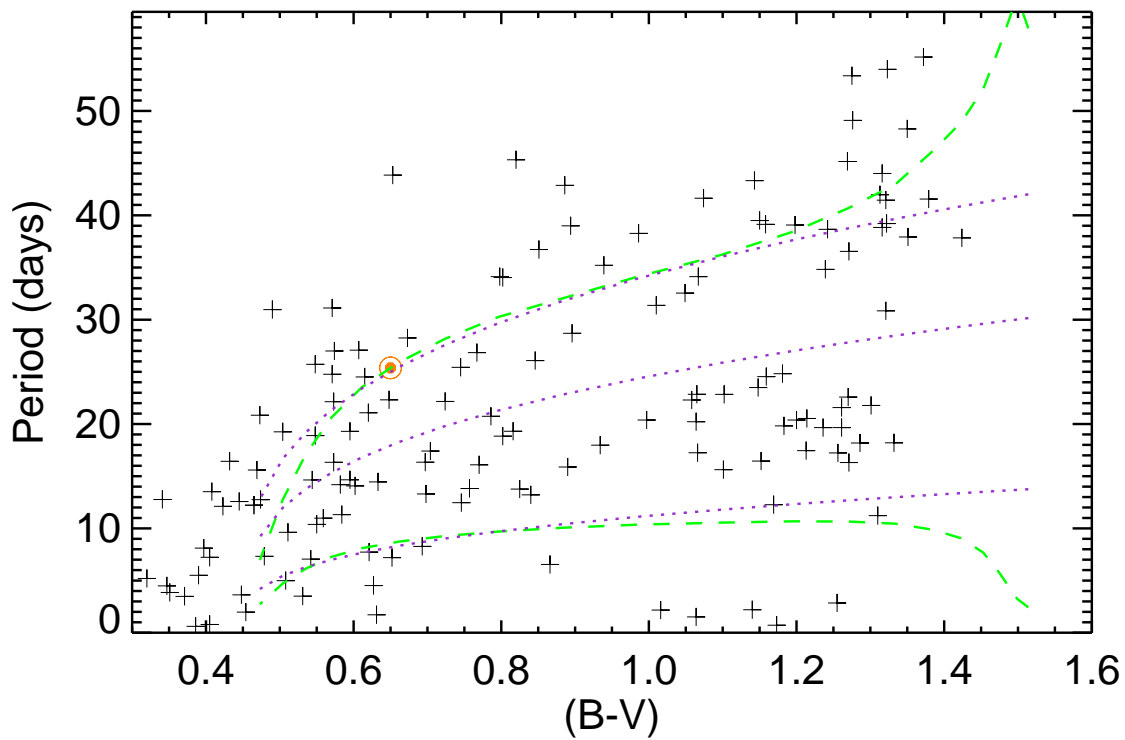}

\caption{Period-color diagram for all active stars with reliable periods.
The Sun's position is shown
with a dotted circle. Period-color isochrones are shown for 4.57 Gyr, 2.5 Gyr 
and 600 Myr stars. The green lines are from \citet{barnes10a} and 
the purple lines are from \citet{angus15}.  The upper two lines are the
4.57 gyr isochrones.  The purple dotted line at intermediate periods 
represents  
a 2.5 Gyr isochrone from \citet{angus15}.
\label{gyro2}}
\end{figure}

Since a number of the stars in the present sample are missing B-V
photometry, it was convenient in sections 3 and 4 to use (g-K) as the
color index.  But to permit comparison with previous work, I have 
taken photometric data compiled in the MAST to transform the
(g-K) colors into (B-V). Figure \ref{gk2bv} shows the relation 
between (g-K) and (B-V) for those
stars in the CPM sample that have both colors.  A simple quadratic
least-squares solution to the data gives:
\begin{equation}
\label{gkbv}
(B-V) = (-0.103 \pm 0.002) + (g-K) (0.424 \pm 0.000) - (0.028 \pm 0.006) 
           (g-K)^2.
\end{equation}
Figure \ref{gyro1} shows the period-color relation for the 56 active CPM
pairs, with B-V values derived from equation \ref{gkbv}.
The solid lines in the figure connect the hotter star (blue symbol) 
of each pair with its cooler companion (red dot).  Figure \ref{gyro2} is
a companion figure showing all 141 active stars without the connecting lines.

For comparison, the Sun's position in these diagrams is shown with a dotted 
circle.  For the age of the Sun, I took the recent meteorite value of 
\citet{connelly12}, 4.57 Gyr.  Because of differential rotation, the Sun 
(presumably along with most of the other stars discussed in this paper) does
not have a unique rotation period.  I took the sidereal Carrington Rotation 
Period value of 25.38 days, which corresponds to the solar rotation at
30\degr\ latitude, the typical latitude of sunspots.  

\subsection{The period-color distribution}

\begin{figure}

\plotone{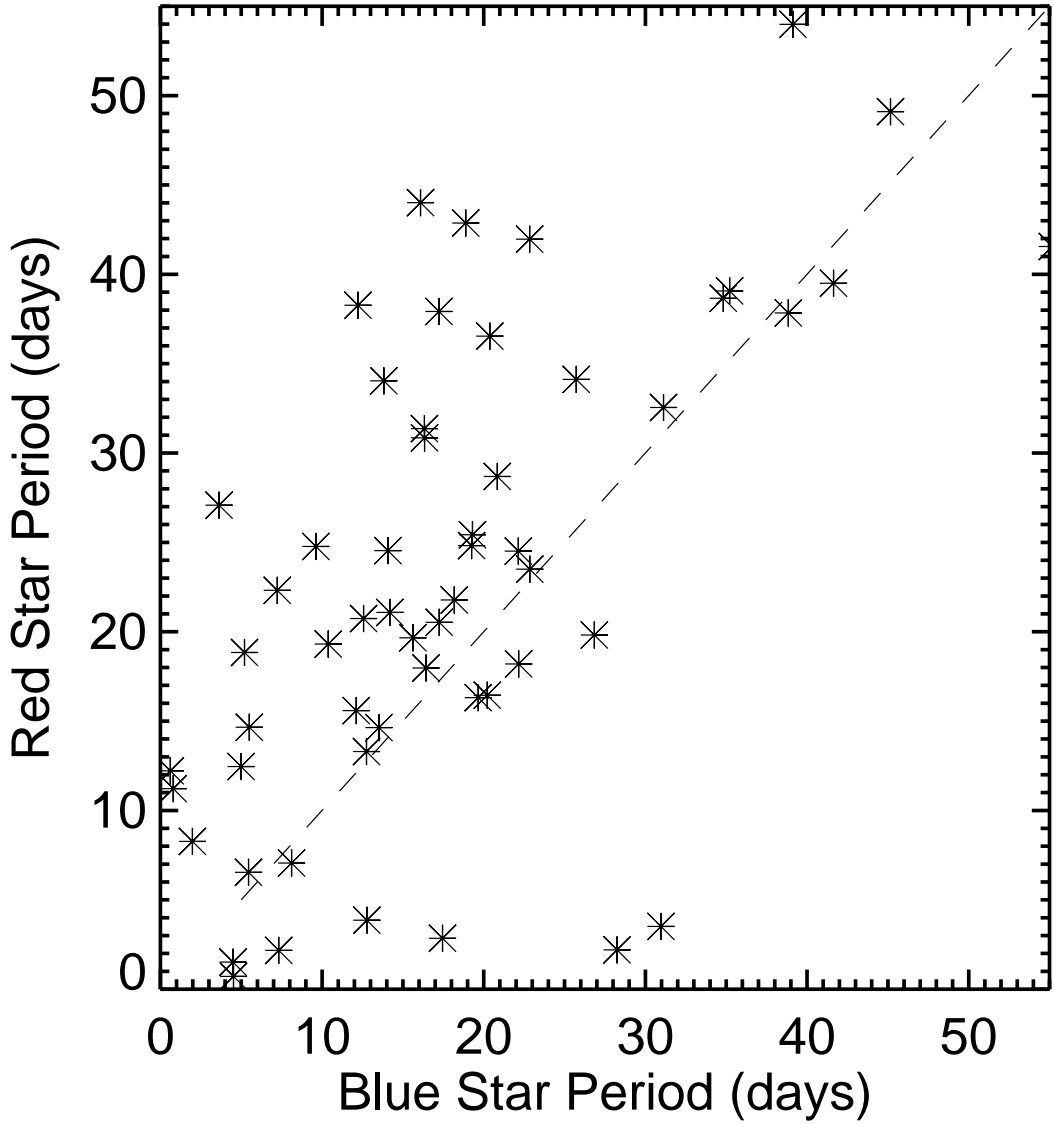}

\caption{Periods of the blue member of each pair vs the red component period.
The solid line in the figure has a slope of one.  Nearly all of the red 
components have longer periods than their blue companions.
\label{pp}}
\end{figure}

There is a general progression to longer periods at cooler 
temperatures in the data of Figure \ref{gyro1} and, as Figure \ref{pp} 
illustrates, nearly all of the
redder stars have longer periods than their bluer companions. The dashed line
in the figure represents the lower envelope of the period-period distribution.
This trend toward longer
periods at cooler temperatures is
consistent with the
various models mentioned above, but there is a wide range in the 
individual gradients, including several pairs with negative slopes.

\begin{figure}
\plotone{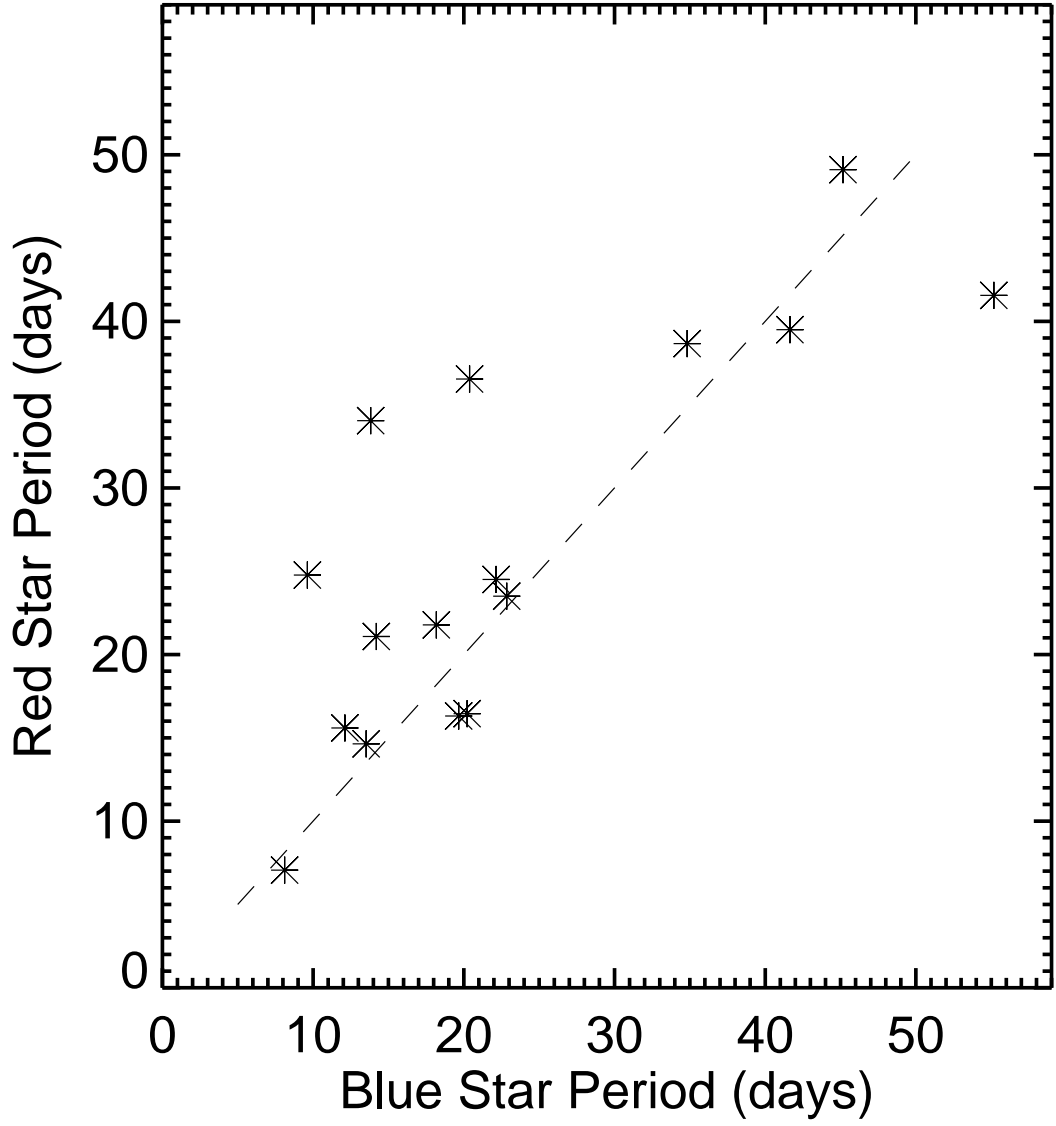}
\caption{Periods of the 2 components of pairs with (g-K) color differences
less than 0.5 magnitudes.  For reference, the line has a slope of one.}
\label{equal}
\end{figure}

If indeed stars of the same mass spin down uniformly with age, 
the rotation periods of the
two (presumably coeval) components of equal mass binaries should be
the same, within the uncertainties.  There are 16 active 
CPM pairs in Table \ref{pairs} with
(g-K) colors within 0.5 magnitudes of one another and 
periods longer than 5.0 days; 
over such a short range in color, the two
components can be expected to have almost
the same mass and so should have nearly 
the same period.  Figure \ref{equal}
shows that although there is an apparent 
correlation between the periods of the A
and B components, there is also considerable scatter. 
(There is also a small bias 
toward longer periods for the red component compared to the blue one, as
expected.) The 
Pearson correlation coefficient of the x and y values of the
points in Figure \ref{equal} is r = 0.80. In 1000 tries of random pairings
of the 32 stars, the correlation coefficient never exceeded this value,
indicating a confidence level in excess of 99.9\%.  
The RMS difference between the red components and blue components is 6.2 days,
much larger than the typical measurement uncertainty in the periods (see Table
\ref{pairs}). So while the periods of the red and blue components are
correlated, in some cases, there are large differences in their periods

\subsection{The age-period-color relation}

\citet{barnes07, mamajek08, angus15} and others
have proposed that rotation periods vs mass (i.e., effective 
temperatures or colors) and
vs ages are separable functions, $P = f(T_{eff}) g(A)$.  Barnes assumed
the \citet{skumanich72} power-law relation for $g(A)$.  \citet{mamajek08} and
\citet{angus15} found slightly different power values for the age factor, 
$P \propto A^{0.56}$\ and $P \propto A^{0.55}$ respectively.  All three of
these groups derived a power law of (B-V) color index to represent effective
temperature (or main sequence mass).  More recently, 
\citet{barnes10a} developed a
more detailed theoretical formulation of the age-period-color relation.
His isochrones (period-color relations at fixed age)
for 4.57 Gyr and 600 Myr stars are also shown in Figure \ref{gyro2}. The 
dashed green lines are from \citet{barnes10a} and the purple dotted lines are 
from \citet{angus15}. The upper two lines are the 4.57 Gyr isochrones. 
The central dotted purple line represents an Angus, et al. 2.5 Gyr 
isochrone. The 
\citet{barnes10a} isochrones are taken from a tabular version of his
Figure 11 which he kindly provided.

\begin{figure}
\plotone{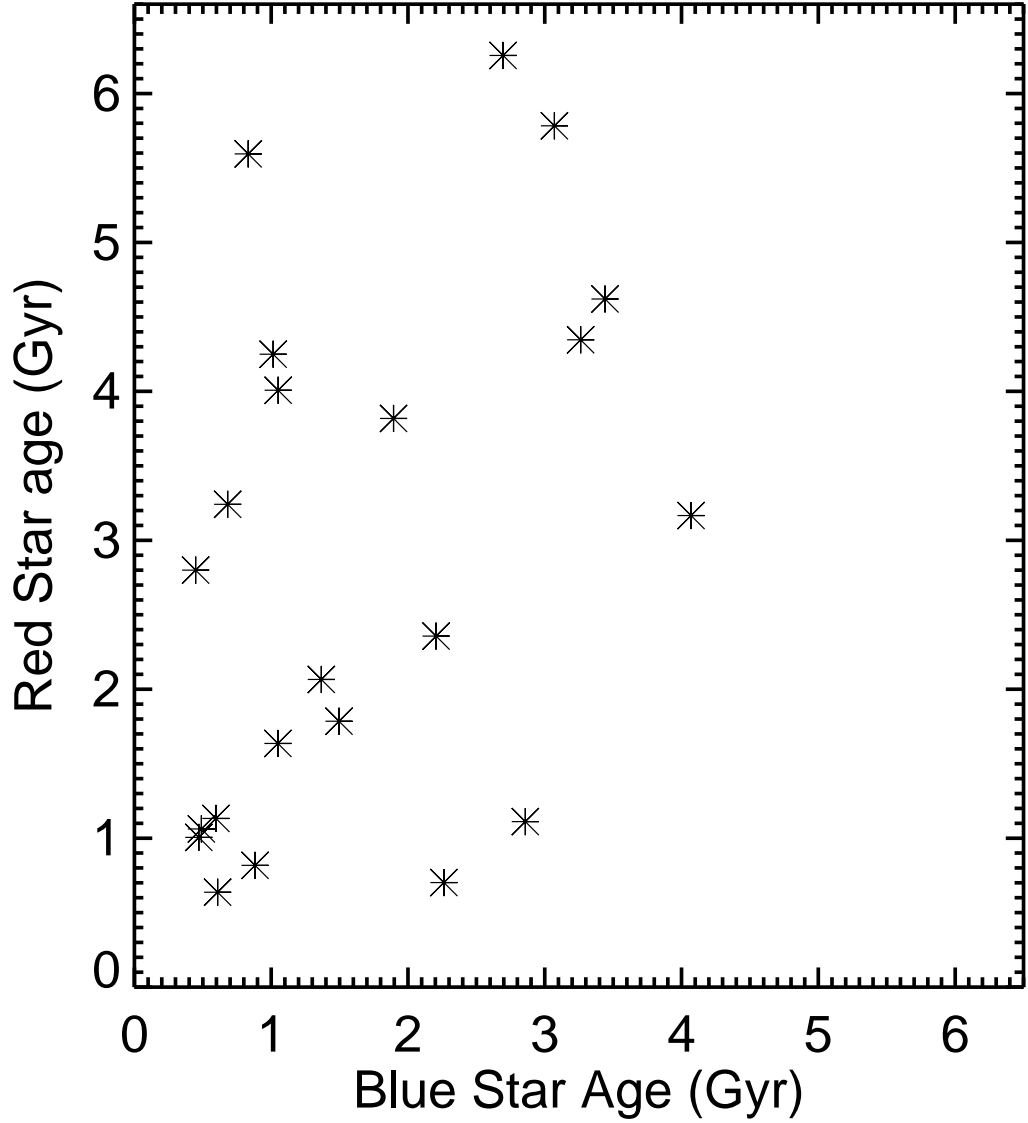}
\caption{Ages of the 2 components of the ``best'' binaries (those with
$(B-V) > 0.6$\ and $P > 5$\ days)
calculated using the \citet{barnes10a} procedure.}
\label{ages}
\end{figure}

Using the \citet{barnes10a} procedure, I 
calculated the ages of the ``best''
selection of systems from the catalog. 
As Figure \ref{gyro2} shows, the gradients
of the period-color curves
of the hotter stars with $(B-V) > 0.6$ are expected to be very steep, 
and the short-period stars with P less than 5 days are either
very young or possibly they have tidally-locked close companions; I
excluded both groups. The 
derived ages are shown in Figure \ref{ages}; there is no obvious correlation
between the calculated ages of the components.

So although Figures \ref{gyro1} and \ref{gyro2} support the gyrochronology 
concept in a general sort of way,
the heterogeneous slopes of the lines connecting the binary pairs 
are not consistent with
a simple age-period-color relation.

\subsection{The upper envelope of the period-color distribution}

Figures \ref{gyro1} and \ref{gyro2} show that near the solar color, there are 
relatively few stars with measureable rotation periods longer than the Sun's
period. 
Furthermore, since the lines connecting the binary components represent lines
of equal age, then the figure as a whole also 
suggests that there may be no more than a
few {\it active} 
stars older than the Sun, even though the Sun is only 
about half the age of the disk.  
This same upper envelope of rotation periods can be seen
in a much larger sample of Kepler stars in Figure 1 of \citet{mcquillan14}. 

Recently \citet{vansaders16} claimed that past the solar age,
weakened magnetic braking results in a decline, or cessation in the rate
of slowing of surface rotation. In their analysis, they made use of 
asteroseismic ages.  But at least in the F, G and early K 
spectral types, stellar activity declines as the star slows down (see Figure
\ref{active}), and older stars (i.e., those with longer periods) might not
be sufficiently active for their rotation periods to be measurable.
Furthermore, 
the wide binary star systems in this sample are weakly bound or not bound at 
all; any older pairs 
may drift far enough apart to be unrecognizable as common proper motion
stars.  So the apparent lack of binary systems with longer periods 
in Figures \ref{gyro1} and \ref{gyro2} cannot
be used to address the \citet{vansaders16} hypothesis. That is to say, if there
are stars at, say the solar color, with longer periods than the Sun, such 
stars would likely not show up in this sample.

\subsection{Gaps in the period distribution}

Finally, beginning about $(g-K) = 3.0$,
there is an apparent gap (visible most clearly in Figure \ref{gyro2})
in the number of stars with intermediate periods
around P = 20 days.  In the present sample, there are also only a few
stars with periods less than about 10 days. 

\citet{barnes07, barnes10a, barnes10b} have
described two stages in the rotational
evolution of solar-type stars. In the initial ``C'' stage, stars are rapidly
rotating but with a range of initial velocities.  They spin down for several
hundred million years until
they converge to the ``I'' 
sequence. By that 
time they have lost all ``memory'' of the initial conditions, and 
thereafter they
spin down according to the \citet{skumanich72} law.  

\begin{figure}
\plotone{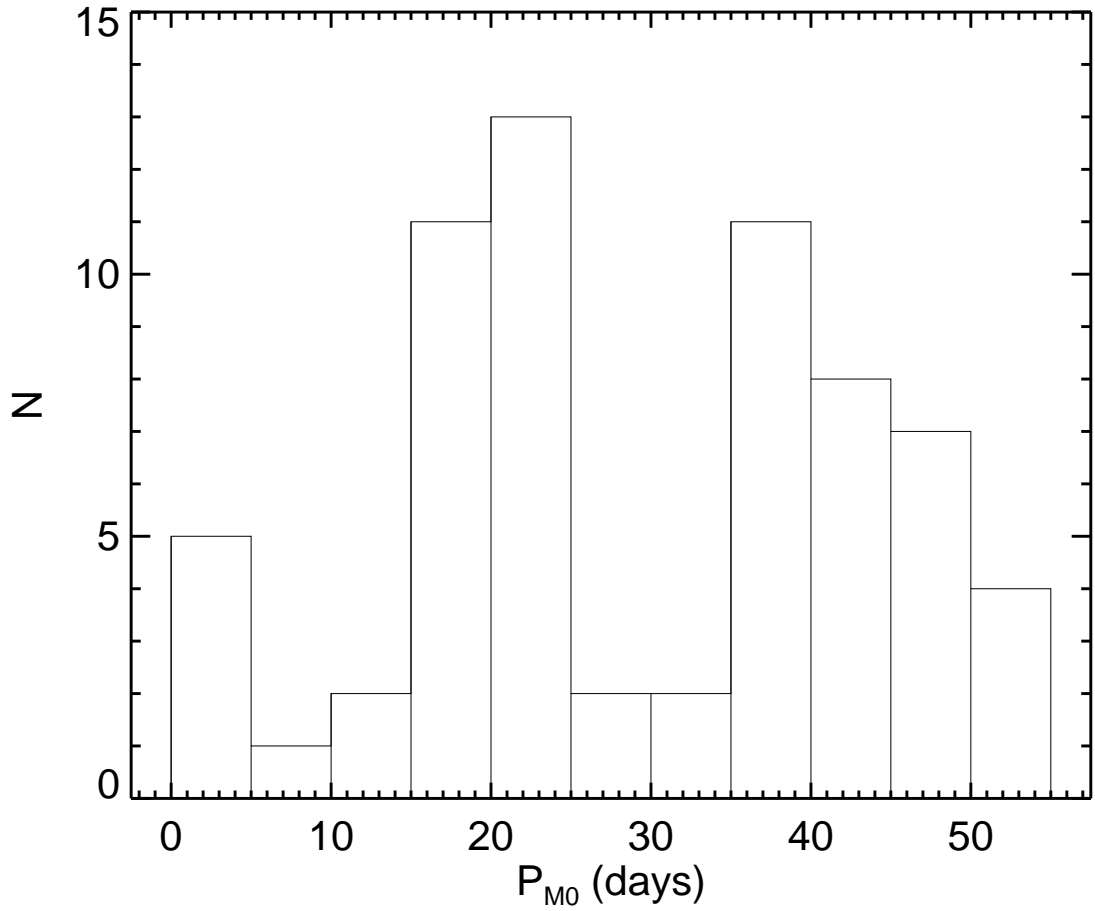}
\caption{Histogram of stellar periods projected to the periods they would have 
at B-V = 1.27 (M0) relative to the period of a star along the 2.5 Gyr isochrone
in Figure \ref{gyro2}.}
\label{phist}
\end{figure}

The figures 
show that while there is a relatively uniform distribution of periods 
among the 
hotter stars, 
a gap is visible in Figure \ref{gyro2} (and somewhat
less obviously in Figure \ref{gyro1}) in the
period
distribution of the cooler stars at intermediate periods.  The gap can be
seen more clearly in Figure \ref{phist}, which shows the period
distribution of active stars 
projected to the periods they would have at (B-V) = 1.27
(about spectral type M0) relative to the period a star on the
2.5 Gyr isochrone would have at that same color.  So for star i,
\begin{equation}
P_{M0}(i) = P(i) [{P_{2.5}(M0) \over P_{2.5}(i)}],
\label{pm0}
\end{equation}
where $P(i)$ is the measured period of star $i$, $P_{2.5}(i)$\ 
is the period of a
star on the 2.5 Gyr isochrone at the color of star $i$, and $P_{2.5}(M0)$\ is
the period of a star on the isochrone at spectral type M0.
Both the ``C/I'' break near 10 days described by \citet{barnes10a} and
the intermediate-period gap near 30 days are visible in this histogram.

\citet{mcquillan13} described {\bf the 30 day} gap in their much larger
sample of late K and early M stars from the Kepler database, and
\citet{newton16} saw a similar gap among mid-M stars in the 
Mearth survey. {\bf \citet{davenport16}, 
after matching a subset of the McQuillan, et al. sample
with the recent \citet{gaia16} data release 1, found evidence for a gap in
the period distribution of bluer stars at periods near the 600 Myr
rotational isochrone. He also traced the 600 Myr isochrone through the
M-star gap, although that is not consistent with the appearance of Figure
\ref{gyro2}.}

\citet{mcquillan13}  {\bf and \citet{davenport16}}
interpreted the gap as being the result of
two distinct waves of star formation. 
However, if that were the case, one might
expect the gap to extend through the entire range of colors (or masses).
\citet{newton16} interpreted the gap in their period data to a non-uniform
rate of evolution, where after evolving slowly for several Gyr, stars enter
a period of more rapid period evolution.
On the other hand, \citet{epstein14} have 
argued that even the long-term evolution of the lowest-mass stars is affected
by the initial conditions. 

As Figure \ref{gyro1} shows, there are several pairs of stars with 
quite different periods, one component being
above the gap and the other below the gap.  This suggests
that the rotational evolution of the late K and early M
stars proceeds in a non-uniform fashion, with some stars evolving in period 
more rapidly than others of the same mass.

\section{Discussion}

Figures \ref{gyro1} and \ref{gyro2} exhibit a
considerable degree of complexity in the rotation period-age-color 
distribution of G, K and M dwarf stars. Some of the apparent 
complexity may be partly a
consequence of the nature of the present sample of stars.  
There are several possible
explanations:

1) Some of the pairs may not in fact be physical
companions.   As I described in Section 2, as many as 15\% of the pairs 
selected on the basis of their proper motions could be accidental. Although 
eight of the original 102 pairs have photometric properties inconsistent with
being main-sequence stars at the same distance and have already been rejected
(see Figure \ref{pdiffs}),
perhaps a few more of the 
56 pairs remaining in 
Figure \ref{gyro1} could also be unrelated to one another.

2) Some ``pairs'' are likely to be hierarchical triple 
systems consisting of a compact, tidally-locked pair and a much wider 
companion.  I removed 
some of the stars with rapid variations earlier
because they are probable close binary systems.  But there could also be some
longer-period stars whose periods are affected by tidal interactions
as well.

3) The apparent starspot period may not be the true rotation
period. For example if 
there are spots on opposite hemispheres, the true period
could be twice the measured value. 
Wherever the autocorrelation function
showed an obvious odd-even pattern of peak heights, I took the double period
\citep[as did][]{mcquillan14}.  But there is also the possibility that even 
when alternate peaks are the same height, the actual rotation period could be
twice the apparent period. 

4) Because of differential rotation, the measured period over a time frame 
even as long as four years may not be the most representative period for a 
star's rotation period.  In the Sun, the mean latitude of the spots, hence
the solar rotation period (as measured by spots) varies over the solar
cycle.  \citet{epstein14} discussed this matter at some length and considered
that differential rotation is a primary limiting factor in determining
the rotational ages of late-type stars.

5) Among the hotter stars in this sample, the steep gradients make the
age calculation nearly indeterminate.  This is true for the open cluster
period-color diagrams which show a large range in period at fixed color at
the blue end of the distribution. (See, e.g., \citet{meibom11}.) A few of the
hotter stars may actually be pulsators, rather than spotted stars.

6) There may not be a single period-mass-age relation.  \citet{barnes10b} 
have argued that after a few hundred million years, a star loses
its ``memory'' of its initial rotation period and stars converge to a single
period at a given age and temperature.  But there may be other circumstances,
such as companion stars or planets, to cause the gyro ages to diverge.
In particular, the rotational evolution of otherwise apparently identical 
K and M stars could proceed in
a non-uniform way, opening gaps or peaks
in the period distribution at fixed color.

It is likely that at some level, all of these possibilities are operating.  
As stars evolve rotationally, their periods will gradually diverge from a mean
period for the particular mass and age.

\section{Summary}

Using existing proper motion catalogs, I was able to identify 93 likely common
proper motion pairs in the Kepler field. I was able to find reliable periods
for about 2/3 of the stars, with 56 complete pairs with periods.  The cooler 
(secondary) companions tend to have longer periods than the warmer (primary)
components.  The overall activity level of the cooler stars  is
at best only a weak function of rotation period.  There are indications
that either the rotation period evolution of the cooler stars proceeds at a
non-uniform rate or some stars evolve more quickly than others of the
same mass.

The great variety in the rotational properties of these stars and the small
sample size make it impossible to derive a comprehensive period-age-color
relation from the current data.
Because of the large proper motion 
uncertainties in the existing catalogs, it was necessary to ignore the
large majority of stars with small proper
motions in searching for common proper motion pairs.  
However, future data releases from the Gaia mission 
(see www.cosmos.esa.int/web/gaia/release) should make a dramatic
change in the number of (wide) visual double stars available for
evaluating the rotational history of the cool stars.  

\acknowledgments

I want to  thank Sydney Barnes for his help in preparing this
paper and for providing me with rotational isochrones for Hyades and 
solar age stars.
This research has been supported by the National Aeronautics and Space 
Administration under grant 13-ADAP13-0196.
This research has made use of the MAST database at the Space Telescope 
Science Center and the 
VizieR catalogue access tool, CDS, 
Strasbourg, France.  The original description of the VizieR service was 
published in 
\aaps, 143, 23.  This publication makes use of the Two-Micron All Sky Survey
which is a joint project of the University of Massachusetts and the Infrared
Processing and Analysis Center/California Institute of Technology, funded by
NASA and the NSF.  I want to acknowledge the constructive comments of an
anonymous  reviewer that led to a number of improvements to the final version.

{\it Facilities:} \facility{Kepler}



\clearpage

\clearpage

\end{document}